\documentclass[fleqn,usenatbib]{mnras}
\usepackage{newtxtext,newtxmath}

\usepackage[T1]{fontenc}
\usepackage{ae,aecompl}
\usepackage{booktabs}

\usepackage{pdflscape}

\usepackage{graphicx}	
\usepackage{amsmath}	

\usepackage{adjustbox}
\usepackage{longtable}
\usepackage{tabularx}
\usepackage{color}  
\usepackage{caption}

\usepackage{tipa}
\usepackage{accents}
\newcommand{\ubar}[1]{\underaccent{\bar}{#1}}

\definecolor{dgreen}{rgb}{0.0,0.6,0.2}

\newcommand{\ie}{$i.e.,$ }
\newcommand{\kms}{km\,s$^{-1}$}

\newcommand{\teff}{T$_{\rm eff}$}
\newcommand{\logg}{log\,$g$}

\newcommand{\msun}{M$_{\odot}$}

\title[Outskirts of UFDs]{The Cosmic Hunt\thanks{The Cosmic Hunt: Variants of a Siberian – North-American Myth, \citet{Berezkin05}.} for Members in the Outskirts of Ultra Faint Dwarf Galaxies: Ursa Major I, Coma Berenices, and Bo{\"o}tes I}

\author[F. Waller et al.]
{Fletcher Waller,$^{1}$\thanks{E-mail: wallerd@uvic.ca, kvenn@uvic.ca}
Kim A. Venn$^{1}$,
Federico Sestito$^{1}$,
Jaclyn Jensen$^{1}$, 
Collin L. Kielty$^{1}$, \and
Asya Borukhovetskaya$^{1}$,
Christian Hayes$^{1,2}$,
Alan W. McConnachie$^{1,2}$, \and
and Julio F. Navarro$^{1}$
\\
$^{1}$Department of Physics and Astronomy, University of Victoria, Victoria, BC, V8W 3P2, Canada \\
$^{2}$ NRC Herzberg, Victoria, BC, Canada \\
}

\date{Accepted XXX. Received YYY; in original form ZZZ}

\pubyear{2022}

\begin{document}
\label{firstpage}
\pagerange{\pageref{firstpage}--\pageref{lastpage}}
\maketitle

\begin{abstract}
Gaia EDR3 data was used to identify potential members in the outskirts of three ultra faint dwarf (UFD) galaxies; Coma Berenices ($>2$R$_h$), Ursa Major I ($\sim4$R$_h$), and Bo{\"o}tes I ($\sim4$R$_h$), as well as a new member in the central region of Ursa Major I. 
These targets were observed with the Gemini GRACES spectrograph, which was used to determine precision radial velocities and metallicities that confirm their associations with the UFD galaxies. 
The spectra were also used to  measure absorption lines for 10 elements (Na, Mg, K, Ca, Sc, Ti, Cr, Fe, Ni, and Ba), which confirm that the chemical abundances of the outermost stars are in good agreement with stars in the central regions.  
The abundance ratios and chemical patterns of  the stars in Coma Berenices are consistent with contributions from SN Ia, which is unusual for its star formation history and in conflict with previous suggestions that this system evolved chemically from a single core collapse supernova event.
The chemistries for all three galaxies are consistent with the outermost stars forming in the central regions, then moving to their current locations through tidal stripping and/or supernova feedback. 
In Bo{\"o}tes~I, however, the lower metallicity and lack of strong carbon enrichment of its outermost stars could also be evidence of a dwarf galaxy merger.
   
\end{abstract}

\begin{keywords}
stars: abundances, Population II -- galaxies: dwarf, evolution, formation -- Local Group
\end{keywords}

\section{Introduction}
\label{intro}

Dwarf galaxies are among the oldest and least chemically evolved objects in the universe \citep[e.g.,][]{Tolstoy2009, Bullock2017}. They include the most dark matter dominated systems known, the ultra-faint dwarf (UFD) galaxies, which are defined as galaxies with absolute magnitudes M$_V>-7.7$ ($L<10^5L_{\odot}$; \citealt{Simon2019}). Their dynamical mass-to-light ratios (M/L) reach 100-1000, and metallicities are less than 1\% solar \citep{McConnachie2012, Simon2019}. Such systems provide interesting challenges and unique opportunities for testing our understanding of dark matter, and for studying the dark energy plus cold dark matter ($\Lambda$CDM) cosmological paradigm.
In addition, recent hydrodynamical simulations of low mass galaxies embedded in dark matter halos have shown that star formation is quenched at very early times, consistent with quenching from cosmic reionization \citep{Wheeler2019, Applebaum2021}. This is similar to the reconstructed star-forming histories found for some UFDs \citep{Brown2014, Bechtol2015}, suggesting that UFDs are possibly the modern day relics of the earliest galaxies. 
The relatively unevolved nature of UFDs make them ideal objects to 
provide insights into galaxy formation and nucleosynthetic events in the early universe.
%
 

Simulations also suggest that the Milky Way (MW) halo has grown from the accretion of these small systems \citep{Bullock2005, Starkenburg2017a, ElBadry18}, providing another way to find and study UFDs -- as disrupted, coherent metal-poor structures, \ie stellar streams.  The exquisite data from the Gaia satellite \citep{GaiaDR2} has uncovered many new structures in the MW halo \citep{Martin2022, Li2022}. 
As UFDs are by definition extremely faint systems, with only a handful of stars that are bright enough for detailed analyses \citep{McConnachie2012, Simon2019, McVenn2020}, chemo-dynamical studies have been hindered by cosmic variance and stochastic sampling. 
For example, combining Gaia DR3 data, spectroscopy, and direct dynamical modelling over a large extent (4.1 half-light radii) in Boo\,I, \citet{Longeard22} has found that this UFD is more elongated than previously thought 
and may have been affected by tides, thereby strengthening links between nearby UFDs and stellar streams in the MW halo.

%
%

Nevertheless, while some of these streams appear indeed to be remnants of accreted dwarf galaxies \citep{Ibata01, Ibata21, Thomas2022, Li2022}, while others are most likely disrupted globular clusters \citep{Malhan2019, Martin22b, Li2022}.
Distinguishing streams that originate from UFDs or globular clusters is greatly helped by detailed chemical abundances from high resolution spectroscopy.  Massive, old globular clusters will show variations in their light elements \citep[CNO, Na, Mg, Al;  ][]{carretta2009anticorrelation, Bastian2018} attributed to contributions from multiple populations of stars.
In UFDs, the location and specific properties of supernovae and compact binary merger events can impact which stars become enriched and by how much \citep{Leaman2012, Nomoto13, Reggiani2017, Kobayashi2020}.  
Thus, the specific abundance ratios of a variety of elements
can provide clues to the characteristics of the initial star formation events, whether streams are disrupted globular clusters or UFDs. 
%


Most UFD stars with detailed analyses are bright 
(V$<$19) and situated close to the projected UFD centres.  Yet—recently several candidate members stars have been found at very large half-light radii (R$_{\rm h}$) (\citealt{McVenn2020, PaceLi2019}), where R$_{\rm h}$ is the radius within which half of the galaxy's light is contained, measured along the semi-major axis. This includes one star in Tuc~II at $\sim$ 9 R$_{\rm h}$, which  demonstrates that Tuc~II is remarkably extended as a result of either strong bursty feedback, an early galactic merger, or tidal interactions with the MW halo \citep{Chiti21}.
Coma Berenices (Com\,Ber) is a similar UFD to Tuc~II, with an old age from stellar isochrones, low metallicity from spectroscopic analyses, and candidate members located at large half light radii, $> 4$R$_{\rm h}$ \citep[e.g.,][]{Frebel2012, Brown2014, McVenn2020}. 
Three stars in Com\,Ber have surprisingly low heavy element abundances; in particular, their s-process 
abundances (Y, Sr, Ba, Ce) are all lower than known stars of similar metallicity in the MW halo ([Fe/H] $\sim -2.5$; \citealt{Frebel10}). 
This unique chemical fingerprint has been interpreted as a result of enrichment by a single metal-poor supernova event ($>20$ M$_\odot$), with little to no later chemical evolution.
This interpretation would make Com\,Ber a relic from early times, where only one stellar generation may have formed after the first, Population III, SN explosions \citep{Frebel2012}.

Other UFDs have shown very different chemical evolutionary patterns.  Spectroscopic studies of stars in
Reticulum II \citep{Ji16, Roederer16} and Tucana III \citep{Hansen2017} have shown a mixture of r-process normal and r-process rich stars, which is interpreted as a result of stochastic sampling and late enrichment from a rare neutron-binary merger \citep[like GW170817;][]{Tanvir2017}. The large variation in abundance patterns intferred from only a handful of stars in each UFD shows why it is necessary to observe as many bright stars in these systems as possible, to overcome cosmic variance effects and to derive further constraints on nucleosynthetic sources, progenitor masses, chemical yields, gas mixing, and stochastic sampling \citep{Su2018, Wheeler2019, Kobayashi2020, Applebaum2021}.
In this way, UFDs can provide unique information on star formation properties and on the origins of the elements at the earliest times. 
 
In this paper, we add to the canon of high-resolution spectral analyses of bright stars associated with UFDs.  We present the analysis of five stars in three UFDs; Coma Berenices (Com\,Ber), Bo\"otes I (Boo\,I), and Ursa Major I (UMa\,I). Targets were selected with high probability memberships but at large half-light radii (see Section~\ref{sec:target}).  Observations and spectral data reduction are described in Section~\ref{sec:obs}.  Stellar parameters for the model atmospheres analysis are discussed in Section~\ref{sec:atmos}.  Spectra were analysed using both spectral line equivalent widths and spectrum synthesis fitting in Section~\ref{sec:spectra}.  Chemical abundances are discussed in Section~\ref{sec:chem} and placed into the context of the evolutionary history of each UFD in Section~\ref{sec:evol}.


\begin{figure*}
	\includegraphics[width=6.9in]{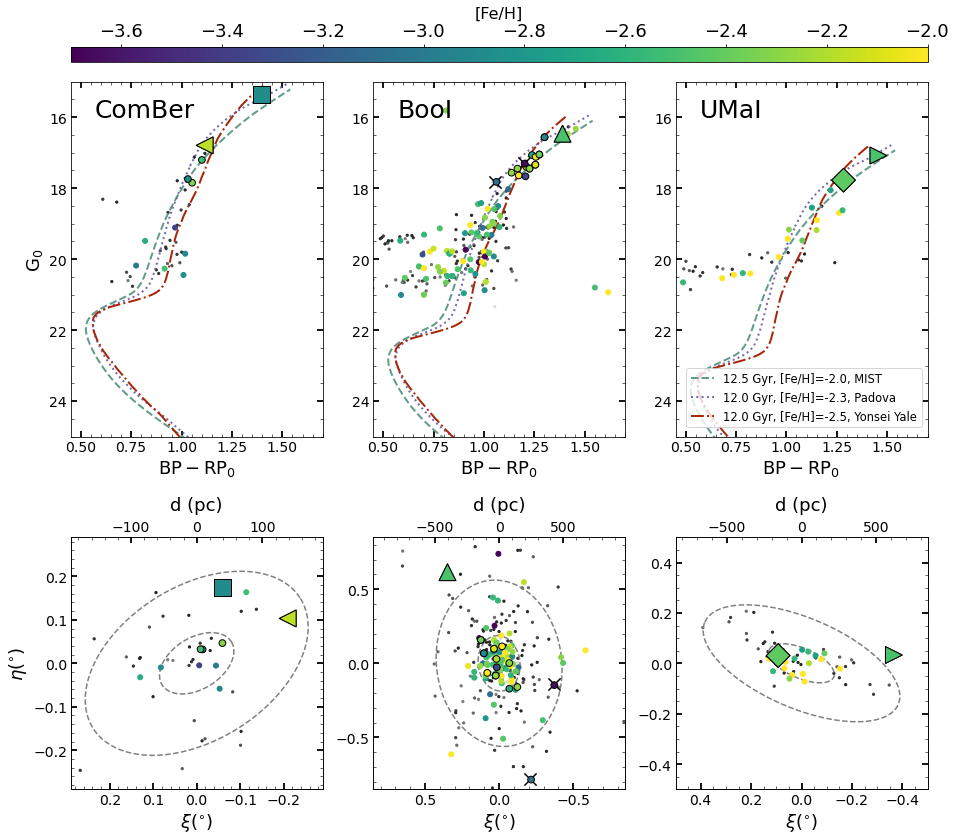}
    \caption{Top row: Gaia colour-magnitude diagram positions of our targets (large symbols) and those in the literature in three UFDs (see Appendix) coloured by their metallicities.  Distances and reddening values used are Table~\ref{tab:obs}. See text for information on the sample isochrones.
    Bottom row: On-sky positions in projected coordinates of stars in these UFDs with spectroscopic metallicities. Dashed lines at 1 and 3 R$_{\rm h}$. Galactic ellipticities, position angles, and half-light radii are listed in Table \ref{tab:galpar}.
    Large coloured symbols are our targets, medium-sized symbols with black outlines are those with high resolution spectroscopy in the literature, and small coloured symbols are those with low resolution metallicities (CaT, Mgb; see Appendix). 
    In Boo\,I, additional stars with HRS analyses at $>2$ R$_{\rm h}$
    are marked with a black ``X" (they are members).
    Targets from Gaia EDR3 with a high likelihood for membership are shown as small dark symbols, grey-scaled weighted by their P$_{\rm sat}$ values from \citet{McVenn2020}. Full list of literature targets is available online.
     }
    \label{fig:halflight}
\end{figure*}

\begin{table*}
\caption{Gaia EDR3 data for each target, as well as our GRACES observational details and their locations in each UFD. In the absence of consistent naming conventions for UFD stars, we use the following naming scheme in this paper, and include Gaia source IDs for cross-matching purposes.  }
\label{tab:obs}
\begin{tabular}{lllccccccccl}
\toprule
 Name  &     RA J2000 &   Dec J2000 & Obs Date   &    G & BP$-$RP & t$_{\rm exp}$ & N$_{\rm exp}$ & SNR  & loc.  & A$_{\rm V}$ & Gaia sourceID \\
 & (hh\,mm\,ss) & (dd\,mm\,ss) & & (mag) & (mag) & (sec) & & (5200, 6000) & (R$_{\rm h}$) & \\
\midrule
CB-1   & 12 26 43.5 & 24 04 45.7 & 2021-01-07 & 15.40 & 1.42 & 900 & 3 & 20, 35 & 2.5 & 0.054 & 3959888486031303424 \\
CB-2   & 12 26 03.9 & 24 00 26.1 & 2021-01-07 & 16.85 & 1.14 & 1800 & 2 & 10, 16 & 2.5 & 0.061 & 3959884740819808256 \\
BooI-2 & 14 01 32.6 & 15 06 50.7 & 2021-05-08 & 16.52 & 1.41 & 2400 & 4 & 15, 30 & 4.0 & 0.058 & 1231264907837100800 \\
UMaI-1 & 10 32 30.1 & 51 57 04.9 & 2021-01-08 & 17.12 & 1.48 & 1800 & 3 & 7, 15 & 3.7 & 0.049 & 847716356845689216 \\
UMaI-2 & 10 35 28.9 & 51 57 01.5 & 2021-01-08 & 17.82 & 1.31 & 2400 & 4 & 7, 15 & 0.7 & 0.052 & 849020961751785472 \\
\bottomrule
\end{tabular}
\end{table*}

\section{Target selection}
\label{sec:target}

Targets were selected using an updated version of a 
new Bayesian inference method for finding highly probable members in UFDs, particularly in extended structures \citep{Jensen2022, McVenn2020}. Gaia photometry and astrometry \citep{Gaia16,Gaia21,Riello21,Lindegren21} was used to assign likelihoods for a star to be a member of one of the UFD galaxies, or of the Milky Way foreground. This was done based on their projected spatial positions and position in color-magnitude space. The models for these likelihoods are based on the derived structural parameters for each dwarf galaxy (including their uncertainties) and the known stellar populations of the dwarfs (including distance uncertainties).  Importantly, radial velocity information is not used when assigning probabilities for individual stars.

A total of ten new targets were found in these three UFDs; six stars in Com\,Ber, three stars in UMa\,I, and one in Boo\,I. We focus primarily on stars with large separations from their galaxy's centers.
In this paper, we present high-resolution spectroscopic (HRS) observations for five bright stars in northern UFDs, Com\,Ber, UMa\,I, and Boo\,I. Their with distances range from central locations to $\sim$ 4R$_{\rm h}$ (see Table~\ref{tab:obs}), where the R$_{\rm h}$ values for each UFD and other structural parameters are in Table~\ref{tab:galpar}. 

This is the first model atmospheres analysis of stars in UMa~I and nearly doubles the number of HRS analyses of stars in Com\,Ber.  We included one star in Boo\,I because of its location (> 3 R$_{\rm h}$).
In the absence of consistent naming conventions for UFD stars, we use the naming scheme given in Table~\ref{tab:obs} in this paper, and include Gaia source IDs for cross-matching purposes.  
The target positions and Gaia colours are listed in Table~\ref{tab:obs} and plotted on the colour-magnitude diagram in Fig.~\ref{fig:halflight}. Reddening values are from \citet{Schlegel1998}.  
%
Sample isochrones are also shown in Fig.~\ref{fig:halflight}, where the
MIST\footnote{MIST/MESA isochrones: \url{http://waps.cfa.harvard.edu/MIST/}} isochrone is from \citet{Paxton2011}, \citet{Choi2016}, and \citet{Dotter2016}, and the Padova\footnote{Padova isochrones: \url{http://stev.oapd.inaf.it/cmd}} isochrone is from \citet{Bressan12}.
The Yonsei-Yale\footnote{Yonsei-Yale isochrones: \url{http://www.astro.yale.edu/demarque/yyiso.html}} isochrone is from \citet{Lejeune1998} and \citet{Demarque04}, where colours were
converted to Gaia photometric bands using the Gaia DR2 Release Documentation (V1.2, 5.3.7). 

Target locations and metallicities are shown in 
Fig.~\ref{fig:halflight} along with metallicities from the literature. For Com\,Ber, these metallicities include medium resolution spectroscopy (MRS, R$\sim$6000) from  \citet{Vargas13} and high resolution spectroscopy (HRS, R$>$30,000) from \citet{Frebel10}.  For UMa\,I, only MRS is available \citep[from][]{Martin07, Vargas13}.
One highly probable member was also found on the outskirts of the southern UFD, Boo\,I.  The metallicities in Fig.~\ref{fig:halflight} for stars in Boo\,I are from both MRS \citep{Martin07, Lai11, Norris10a} and HRS \citep{Gilmore13, Ishigaki14, Feltzing09, Frebel16}.

\tabcolsep=0.17cm
\begin{table}
\caption[]{Galactic parameters for our three UFD galaxies from \citet[][updated 2021\footnotemark]{McConnachie2012} 
RV is the range in radial velocities, D$_\odot$ is the heliocentric distance, {\sl ell} and $\phi$ are the ellipticity and position angle of the isophotes, and R$_{\rm h}$ is the half light radius, for each UFD. }
\begin{tabular}{lrccccc}
\toprule
 Name   & RV range              & D$_\odot$ & ell   & $\phi$      & R$_{\rm h}$ & R$_{\rm h}$\\
        &  (\kms)               & (kpc)     &       & (deg)       & (arcmin) & (pc)\\
\midrule
Com~Ber  &    97.2 to \,\, 99.0 & 40 $\pm$4 & 0.37  &    $-$58    & \ 5.63  &  65 \\
UMa~I    & $-$56.7 to   $-$53.9 & 97 $\pm$4 & 0.57  &   \ \ 67    & \ 8.34  & 235 \\
Boo~I    &    96.9 to     101.1 & 66 $\pm$2 & 0.25  &  \ \ \ 7    &  11.26  & 217 \\ 
\bottomrule
\end{tabular}
\label{tab:galpar}
\end{table}

\footnotetext{Available at: \url{http://www.astro.uvic.ca/~alan/Nearby_Dwarf_Database_files/NearbyGalaxies_Jan2021_PUBLIC.fits} \label{footalan}}
 
\begin{table}
\caption{The stellar parameters, effective temperature (\teff) and surface gravity (\logg) for our \textsc{MARCS} model atmospheres.  Metallicities ([Fe/H]$_{\rm LTE}$) are from our spectral line analysis (see Section~\ref{sec:spectra}), and microturbulence ($\xi$) is calculated from \citet{Mashonkina17}. Values of \logg$<$0.5 were set to 0.5 to avoid extrapolating outside the \textsc{MARCS} grid.  Radial velocities (RV) are found with \textsc{IRAF}/{\sl fxcor} from our spectra. Parameters for HD122563 are from \citet{Mashonkina17} and \citet{Karicova2020}. 
%
}
\tabcolsep=0.08cm
\begin{tabular}{lccccc}
\toprule
Name  & \teff  & \logg  & $\xi$  & [Fe/H]$_{\rm LTE}$ & RV  \\
   &  (K) &  (dex) & (\kms) & (dex) & (\kms) \\
\midrule
CB-1   & 4385 $\pm$81 & 0.33 $\pm$0.12 & 2.1$\pm$0.1  & $-2.88 \pm$0.15   & \,\, 91.76 $\pm$0.18 \\
CB-2   & 4879 $\pm$99 & 1.24 $\pm$0.12 & 2.5$\pm$0.1  & $-2.16 \pm$0.12   & \,\, 95.86 $\pm$0.27 \\
BooI-2 & 4390 $\pm$70 & 0.41 $\pm$0.08 & 2.3$\pm$0.1  & $-2.49 \pm$0.14   & \,\, 99.25 $\pm$1.09 \\ 
UMaI-1 & 4308 $\pm$68 & 0.28 $\pm$0.12 & 2.1$\pm$0.1  & $-2.50 \pm$0.12   & $-$51.73 $\pm$0.38 \\
UMaI-2 & 4574 $\pm$83 & 0.75 $\pm$0.11 & 2.4$\pm$0.1  & $-2.42 \pm$0.20   & $-$53.78 $\pm$0.98 \\
HD122563 & 4635 $\pm$34 & 1.40 $\pm$0.04 & 1.7~~~~~~~~ & $-2.75 \pm$0.12  & — \\
\bottomrule
\end{tabular}
\label{tab:param}
\end{table}

\section{GRACES Observations and Reduction}
\label{sec:obs}

High-resolution spectra have been collected using the Gemini Remote Access to CFHT ESPaDOnS Spectrograph 
\citep[GRACES,][]{Chene14,Pazder14}.
In the 2-fibre (object+sky) mode, spectra are obtained with resolution R$\sim$65,000; however, light below $\sim$4800 \AA\ is severely limited by poor transmission through the optical fibre link. 

\begin{figure*}
\begin{center}
    \setlength{\unitlength}{1cm}
    \includegraphics[width=17.7cm]{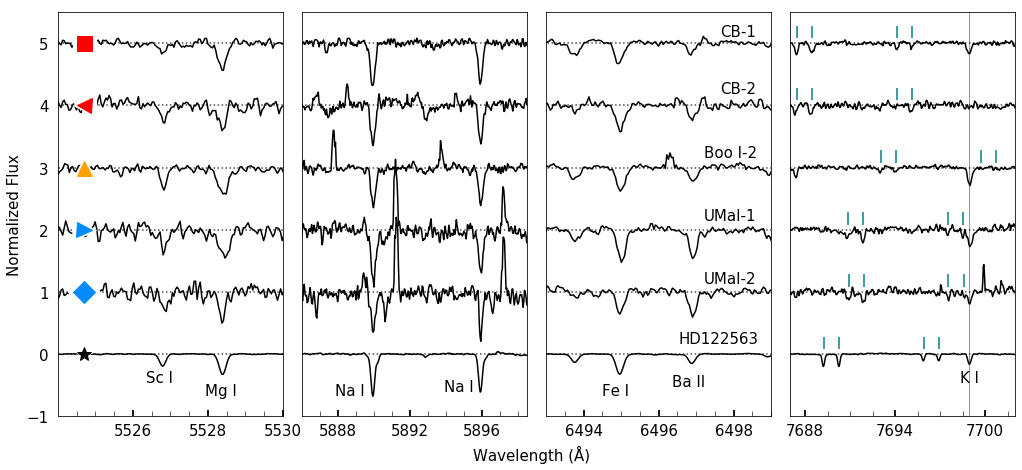}
    \caption{Samples of the GRACES spectra for five stars in three UFD galaxies, and the standard star HD122563. Coloured markers are the same as in Fig.~\ref{fig:halflight}.  The improved SNR at redder wavelengths that is typical of GRACES spectra can clearly be seen in all cases. Key absorption features are marked for clarity. Equally-spaced absorption features from earth's atmosphere are marked in the right hand panel with blue dashes.
    }
    \label{fig:spec}
\end{center}
\end{figure*}

The GRACES spectra have been reduced using the Gemini ``Open-source Pipeline for ESPaDOnS Reduction and Analysis" tool \citep[OPERA,][]{OPERA}, as described in \citet{Kielty2021}.  Briefly, this includes standard calibrations (including wavelength calibration and heliocentric corrections).  Starting from the individually extracted and normalized {\'e}chelle orders, one continuous spectrum is created by weighting the overlapping wavelength regions by their error spectrum, and co-adding as a weighted average.  No radial velocity variations were found between the multiple visits ($\Delta$RV$\le1$\,\kms) for each star, and all visits per star have then been co-added via a weighted mean using the error spectrum.  Each co-added spectrum was radial velocity corrected by comparisons with the metal-poor benchmark star HD~122563, and re-normalized using asymmetric k sigma-clipping. This normalization routine can occasionally under-estimate the continuum due to small absorption lines being hidden in the noise. This is discussed in Section \ref{sec:spectra}.   Sample spectra are shown in Fig.~\ref{fig:spec}.

\section{Stellar Parameters}
\label{sec:atmos}

All five stars in this paper are expected to be red giants based on their magnitudes and membership in each UFD, as shown in Fig \ref{fig:halflight}.  This affects how stellar parameters are determined, which is described below.  Our final stellar parameters per star are listed in Table~\ref{tab:param}.


Effective temperatures (\teff) have been determined using the colour-temperature relation for Gaia photometry from  \citet[][hereafter MB2020]{Mucciarelli2020}. This calibration was selected based on their inclusion of very metal-poor stars (from \citealt{GonzalezHernandez2009}). When calculating temperatures from MB2020, it is necessary to know if the star is a dwarf or giant (two sets of calibrations) and to have a metallicity estimate a priori. We adopted [Fe/H]=$-2.5$ for all stars, representative of the metallicity estimates of these UFDs in the literature.  We adopted an uncertainty of $\Delta$[Fe/H]=$\pm$0.5 to propagate into our stellar parameter error estimates. 
A comparison of the temperatures from the MB2020 calibration to those from the colour-temperature relationship by \citet{Casagrande2020} showed very good agreement in metal-poor halo stars \citep[e.g.,][]{Kielty2021}. 


Surface gravities (log~$g$) were determined using the Stefan-Boltzmann equation (e.g., \citealt{Venn17}; \citealt{Kraft2003}). This method requires (i) a distance, which we took from the dwarf galaxy table in \citet[][updated 2021\textsuperscript{\ref{footalan}}]{McConnachie2012}, (ii) the solar bolometric magnitude of M$_{\rm bol}$ = 4.74, and (iii) bolometric corrections for Gaia DR2 photometry \citep{Andrae2018}. Any gravities with log\,$g<$ 0.5 were rounded up to 0.5 to keep them within the model atmosphere grid. 


The metallicities listed in Table~\ref{tab:param} are the final metallicities from our spectral lines analysis described below (see Section~\ref{sec:spectra}).  
Microturbulence is not needed for our stellar parameter determinations; however, we have included it in Table~\ref{tab:param} for completeness as it is required for our model atmospheres analysis in Section~\ref{sec:spectra}. Microturbulence ($\xi$) was adopted from the calibrations for metal-poor red giants by \citet{Mashonkina17}.



\subsection{Stellar Parameters: Uncertainties}
\label{subsec:param_other}

Uncertainties in \teff \, and \logg \, were determined from a Monte Carlo analysis on the uncertainties from the input parameters, as well as assuming a flat prior on the stellar mass, spanning 0.5 to 1.0 \msun.  The stellar parameters used in generating model atmospheres, \teff\ and log\,$g$, and their uncertainties, are listed in Table~\ref{tab:param}.

Uncertainties in metallicity are from our iron spectral line analyses (below). Meanwhile, the uncertainties in the microturbulence ($\xi$) values are initially taken from the calibration by \citet{Mashonkina17}.  For weak lines in our spectral analysis (EW $<150$ m\AA), the uncertainties in $\xi$ have very small to negligible impact on the stellar abundances. However, for a few elements (Mg, Na), we have been forced to keep stronger lines (up to 300 m\AA) since only one or no other spectral lines are available.  Stronger lines are much more sensitive to $\xi$, 
however in all cases, we find the chemical abundance uncertainties are dominated by the uncertainties in \teff\ and \logg, rather than $\xi$; see Section~\ref{sec:errors}.

A spectroscopic method of finding the stellar parameters was also explored for one star with the most lines, by altering the  \teff\ to minimize correlation between abundance and excitation potential, and by altering \logg\ to produce similar abundances for first and second iron excitation states. The best fit \teff\ and \logg\ were found to be within 1$\sigma$ of the parameters found through the MB2020 methods, confirming the quality of our method for determining the stellar parameters.

\section{Spectral Lines Analysis}
\label{sec:spectra}

Radial velocities and chemical abundances are determined for each star from individual spectral lines.  Spectral lines were selected from the recent GRACES analysis of metal-poor halo stars by \cite{Kielty2021}, see Table~\ref{tab:linelist}.
Radial velocities were determined using IRAF/{\sl fxcor}.  Each final combined spectrum was cross-correlated with a GRACES spectrum of the standard star HD\,122563 (from \citealt{Kielty2021}) in the spectral region from 4800 to 6700 \AA. The RV precision was dependent on the SNR of the spectra and stellar parameters of each star. 

Chemical abundances were determined using the stellar parameters discussed above for a classical model atmospheres analysis. These stellar parameters were used to generate spherical model atmospheres with the (most up-to-date) models on the MARCS website \citep{Gustafsson08}. The OSMARCS spherical models were used given that all the stars have log$g$< 3.5.  
Atomic data was adopted from the {\sl linemake}\footnote{ Available at \url{https://github.com/vmplacco/linemake}} atomic and molecular line database \citep{Placco21}.  Isotopic and hyperfine structure corrections were included for lines of \ion{Ba}{II}.
%
Chemical abundances are compared to the Sun using standard notation [X/Y] = log n(X)/n(Y)$_*$ $-$ log n(X)/n(Y)$_\odot$, where n(X) and n(Y) are column densities (in cm$^{-2}$).  We adopt the solar abundances from \citet{Asplund09}.  

Two methods were explored in our chemical abundance analysis: (i) spectral line syntheses, and (ii) line-by-line equivalent width (EW) results. 
%
%
We found that an iterative process between the spectrum syntheses and the EW analysis per line yielded the most reliable results as the signal-to-noise (SNR) of our final combined 1D spectra was typically quite low.  Each of these methods are discussed in the following two subsections.

\begin{table*}
\caption{Sample line list, equivalent width (EW) measurements, and chemical abundances (where X = 12+log(X/H)) from both the EW (X$_{\rm EW}$) and spectrum synthesis (X$_{\rm SYN}$) analyses. Flags are as follows: ``n'' too noisy, ``w'' too weak, ``b'' a blend, ``s'' too strong (EW$>150$ m$\Angstrom$). 
Results in italics are strong lines which were used when no other lines are available. SNR values provided at 5200~$\Angstrom$ per star.
Full line list available online.}
\begin{tabular}{llrrrrrrrrrrrrrrr}
\hline
           &       &       \multicolumn{3}{c}{CB-1 (SNR=20)}   &      \multicolumn{3}{c}{CB-2 (SNR=10)}    & \multicolumn{3}{c}{UMaI-1 (SNR=7)}        & \multicolumn{3}{c}{UMaI-2 (SNR=7)}    &  \multicolumn{3}{c}{BooI-2 (SNR=15 )}   \\
 \cmidrule(lr){3-5}\cmidrule(lr){6-8} \cmidrule(lr){9-11}\cmidrule(lr){12-14} \cmidrule(lr){15-17} 
 Wave. & Elem & EW   & X$_{\rm EW}$ & X$_{\rm SYN}$ & EW   & X$_{\rm EW}$ & X$_{\rm SYN}$ & EW   & X$_{\rm EW}$ & X$_{\rm SYN}$ & EW   & X$_{\rm EW}$ & X$_{\rm SYN}$ & EW   & X$_{\rm EW}$ & X$_{\rm SYN}$ \\
\hline\hline
   5216.274 &    26.0 &  125 &       4.98 &        4.75 &   s  &          — &           — &  n   &          — &           — &  140 &       5.27 &        5.60  &  n   &     —      &        —    \\
   5232.940 &    26.0 &  105 &       4.69 &        4.65 &  140 &       5.52 &        5.05 &  s   &          — &           — &  130 &       5.02 &        5.17  & 115  &   4.64     &      4.98   \\
   5247.050 &    26.0 &   45 &       4.48 &        4.80 &   n  &          — &           — &  130 &       5.08 &        5.40 &  105 &       5.55 &        5.40  &  80  &   4.79     &      4.95   \\
   5250.210 &    26.0 &   80 &       4.91 &        4.70 &   n  &          — &           — &  125 &       5.13 &        4.85 &   90 &       5.34 &        5.68  &  85  &   4.83     &      5.00   \\
 ... \\
\bottomrule
\end{tabular}
\label{tab:linelist}
\end{table*}

\subsection{Spectrum Syntheses}
\label{syntheses}

The 1D LTE radiative transfer code MOOG\footnote{Available at \url{http://www.as.utexas.edu/~chris/moog.html}} \citep[][]{Sneden73, Sobeck11} was used to synthesize the stellar spectra using the stellar parameters as described above.
This method was carried out in three steps: (1) a model atmosphere was generated with the initial parameters.  The iron lines were examined for a preliminary metallicity estimate, and the model atmosphere updated with the new metallicity.  This process was repeated until convergence (typically only twice). (2) A new synthesis of all elements was generated including line abundances and upper limits for all of the clean spectral lines. (3) Hyperfine structure corrections were applied to barium from a full spectrum synthesis within $\pm$10$\Angstrom$ of the \ion{Ba}{II} lines. 

Each synthetic spectra was broadened in MOOG to match the observed spectra; we found that a Gaussian smoothing kernel with FWHM=0.2 was a good match to the GRACES spectral resolution (which dominates the broadening for these low gravity red giant stars). If a spectral feature was well fit, then we calculated an abundance for that line.  If not, then a 3$\sigma$ upper limit was calculated.  Amongst the upper limits that were calculated, none other than one \ion{K}{I} line provided a useful scientific constraint.

Additional spectral lines were sought once we had preliminary results for each of the five targets in this paper.  Any new spectral features found and measured were collated into our master line list (see Table~\ref{tab:linelist}), which was then used for a full  synthesis of each stellar spectrum. 
%
%
The full and final synthesis was carried out with the final stellar and spectral parameters and compared to the equivalent widths method as described below.

\subsection{Equivalent Widths Analysis}
\label{EW}

Equivalent widths (EW) were measured in IRAF/{\sl splot} \citep{Tody86, Tody93} and provided in Table~\ref{tab:linelist}. Weak lines were measured by fitting a Gaussian profile, and stronger lines were examined by also integrating under the continuum or where noise made fitting impossible. 
The continuum was carefully examined, and adjustments upwards by $\le$10\% was occasionally necessary to account for weak lines in the low SNR spectra. 


Abundances were determined by assuming local thermodynamic equilibrium (LTE) and examining the EW line measurements in MOOG. The microturbulence ($\xi$ in \kms) was examined and slightly adjusted to minimize any EW-abundance correlations in the \ion{Fe}{I} lines. Lines with abundances well above or below the 1$\sigma$ line scatter were confirmed to be blends or noise spikes respectively, and discarded. 
Comparison of the equivalent width and the syntheses methods allowed us to remove or flag line-confounding noise spikes (e.g. a noise spike in the middle of the trough).
Iron lines with $EW>$150 m$\Angstrom$ were flagged as strong and discarded.  The lower limit of $EW\sim$30 m$\Angstrom$ was adopted given the SNR of our spectra.  


Correlations between the \ion{Fe}{I} line abundances with excitation potential, equivalent width, and wavelength were confirmed to be minimal before proceeding. This was to check the stellar parameters and data reductions.  Ionization balance between \ion{Fe}{I} and \ion{Fe}{II} was also examined as a check on the surface gravities (including NLTE corrections; see below).  
For three stars (CB-1, UMaI-1, and BooI-2), we were forced to adopt a slightly larger surface gravity (\logg=0.5) which would also have a small effect on the \ion{Fe}{I} and \ion{Fe}{II} abundances. 

The final calculations included all updates from these tests (line lists, metallicities and microturbulence).   As a final step,
hyperfine-structure corrections were applied to the \ion{Ba}{II} analyses.

\begin{table}
\caption{Sample uncertainties for the CB-1 and UMaI-1 analyses (the two stars with the highest and lowest SNR, respectively).  
Systematic errors in the abundances due to the uncertainties in the stellar parameters, \teff, \logg, metallicity and microturbulence from Table~\ref{tab:param} are listed as 
$\Delta T$,   $\Delta g$,   $\Delta m$, and $\Delta \xi$.
The line scatter for the equivalent widths and syntheses methods from Table~\ref{tab:linelist} are denoted as $\sigma_{\rm EW}$ and $\sigma_{\rm syn}$. For species with fewer than 5 lines, the scatter from \ion{Fe}{I} was adopted, reduced by $\sqrt{n}$ of the species. $\Delta EW$ and  $\Delta syn$ are the full errors per element species from the line scatter and the stellar parameter abundance errors, added in quadrature.  Full table for the other four stars in the Appendix.}
\tabcolsep=0.16cm
\begin{tabular}{lrrrrrrrr}
\toprule
\textrm{species} & $\Delta T$ &   $\Delta g$ &    $\Delta m$  &   $\Delta \xi$ &  $\sigma_{EW}$ & $\sigma_{syn}$ &  $\Delta EW$ &  $\Delta syn$ \\
\midrule
\textbf{CB-1}    &       &       &       &       &        &           &             &       \\
\textrm{\ion{Fe}{I}   } &  0.14 &  0.00 &  0.04 &  0.01 &   0.02 &      0.05 &        0.15 &       0.15 \\
\textrm{\ion{Fe}{II}  } &  0.04 &  0.02 &  0.01 &  0.01 &   0.10 &      0.14 &        0.11 &       0.15 \\
\textrm{\ion{Na}{I}   } &  0.18 &  0.01 &  0.06 &  0.03 &   0.14 &      0.17 &        0.24 &       0.26 \\
\textrm{\ion{Mg}{I}   } &  0.10 &  0.04 &  0.03 &  0.02 &   0.11 &      0.13 &        0.16 &       0.17 \\
\textrm{\ion{K}{I}}     &  0.11 &  0.01 &  0.03 &  0.01 &   0.20 &      —    &        0.23 &       —    \\
\textrm{\ion{Ca}{I}   } &  0.08 &  0.02 &  0.04 &  0.01 &   0.03 &      0.08 &        0.10 &       0.12 \\
\textrm{\ion{Sc}{I}   } &  0.12 &  0.01 &  0.02 &  0.02 &  —     &      0.20 &          —  &       0.24 \\
\textrm{\ion{Ti}{I}   } &  0.19 &  0.00 &  0.06 &  0.01 &   0.08 &      0.09 &        0.21 &       0.22 \\
\textrm{\ion{Ti}{II}  } &  0.00 &  0.03 &  0.01 &  0.01 &   0.09 &      0.09 &        0.09 &       0.10 \\
\textrm{\ion{Cr}{I}   } &  0.16 &  0.00 &  0.06 &  0.02 &   0.11 &      0.09 &        0.21 &       0.19 \\
\textrm{\ion{Ni}{I}   } &  0.14 &  0.00 &  0.04 &  0.01 &   0.14 &      0.12 &        0.20 &       0.19 \\
\textrm{\ion{Ba}{II}  } &  0.05 &  0.04 &  0.02 &  0.01 &   0.14 &      0.13 &        0.15 &       0.14 \\
\bottomrule
\end{tabular}
\label{tab:errors}
\end{table}

\begin{table*}
\caption{Sample of the NLTE corrections ($\Delta_{\rm NLTE}$) from two databases, INSPECT and MPIA (see text), such that X$_{\rm NLTE}$ = X$_{\rm LTE}$ + $\Delta_{\rm NLTE}$.  Full table of NLTE corrections is available online. K and Ba corrections are from \citet{Andrievsky10} and \citet{Mashonkina19} respectively, and marked in the Inspect column with a * in the full table. }

\begin{tabular}{lrrrrrrrrrrr}
&        &      \multicolumn{2}{c}{CB-1 }   &       \multicolumn{2}{c}{CB-2 }    & \multicolumn{2}{c}{UMaI-1 }        & \multicolumn{2}{c}{UMaI-2 }    & \multicolumn{2}{c}{BooI-2 }    \\
 \cmidrule(lr){3-4}\cmidrule(lr){5-6} \cmidrule(lr){7-8}\cmidrule(lr){9-10}\cmidrule(lr){11-12}
 Wave. & Elem &  Inspect &  MPIA &  Inspect &  MPIA &  Inspect &  MPIA &  Inspect &  MPIA &  Inspect &  MPIA \\
\midrule
   5216.274 &  26.0 &     0.147 &  0.205 &        —  &      — &         — &     —  &     0.185 &  0.155 &     —     &     —  \\
   5232.940 &  26.0 &     0.074 &  0.236 &     0.062 &  0.057 &         — &     —  &     0.099 &  0.107 &   0.064   &  0.127 \\
   5247.050 &  26.0 &     0.130 &  0.198 &        —  &      — &     0.159 &  0.138 &     0.174 &  0.170 &   0.159   &  0.147 \\
   5250.210 &  26.0 &     0.145 &  0.094 &        —  &      — &     0.152 &  0.064 &     0.173 &  0.084 &     —     &  0.069 \\
   ... \\
\bottomrule
\end{tabular}
\label{tab:nltestub}
\end{table*}

\subsection{Measurement and Parameter Uncertainties}
\label{sec:errors}

The measurement errors from both the synthesis and EW methods are from the line-to-line scatter in the abundances, $\sigma_{\rm syn}$ and $\sigma_{\rm EW}$, respectively (see Table~\ref{tab:errors}).  These represent errors in the continuum placement, local SNR, and atomic data quality.  
For elements with few lines (<5), the standard error from \ion{Fe}{I} was adopted and reduced by $\sqrt{n}$, where $n$ is the number of lines available.

The chemical abundances are also subject to systematic errors from the uncertainties in our model parameters.  For each program star,  we varied \teff, \logg, [M/H], and $\xi$, one at a time, by the uncertainties given in Table~\ref{tab:param}. The corresponding abundance uncertainties are listed as $\delta T$, $\delta g$, $\delta m$, and $\delta \xi$. 

Total abundance uncertainties, $\Delta_{syn}$ and $\Delta_{EW}$, were calculated by adding in quadrature the line-to-line abundance scatter, with the uncertainties imposed by the stellar parameter errors. These are adopted in the abundance plots in Section~\ref{sec:chem}.

\begin{table*}
\caption{Scaled-solar chemical abundances from our EW analyses, except for species labelled with an * which are from syntheses, with and without NLTE corrections (in Table~\ref{tab:nltestub}). Metallicity [Fe/H] is a weighted mean of [\ion{Fe}{I}/H] and [\ion{Fe}{II}/H]. Solar abundances as 12+log(X/H) from \citet{Asplund09}, and [X/Fe] = log(X/Fe)$_*$ $-$ log(X/Fe)$_\odot$. Species such as \ion{Ni}{I} which have no available NLTE corrections still show an altered value in the NLTE table, which is strictly due to the \ion{Fe}{} NLTE corrections.}
\begin{tabular}{lrrrrrr}
\toprule
                          & Solar&                   CB-1 &                     CB-2 &                  UMaI-1 &                 UMaI-2 &                   BooI-2 \\
                     &  log(X/H) &                [X/Fe]  &            [X/Fe]        &         [X/Fe]          &                [X/Fe]  &           [X/Fe]         \\
\midrule
\textbf{LTE} \\
\textrm{[Fe/H]}           & 7.50 &  -2.88$\pm$0.15    (77) &  -2.17$\pm$0.12    (46) &  -2.49$\pm$0.12    (44) &  -2.42$\pm$0.20    (45) &  -2.49$\pm$0.14    (61) \\
\textrm{[\ion{Fe}{I}/H]}  & 7.50 &  -2.88$\pm$0.15    (73) &  -2.15$\pm$0.12    (41) &  -2.51$\pm$0.12    (42) &  -2.42$\pm$0.20    (43) &  -2.49$\pm$0.14    (55) \\
\textrm{[\ion{Fe}{II}/H]} & 7.50 &  -2.97$\pm$0.11 \ \ (4) &  -2.32$\pm$0.11 \ \ (5) &  -2.04$\pm$0.15 \ \ (2) &  -2.46$\pm$0.17 \ \ (2) &  -2.45$\pm$0.10 \ \ (6) \\
\textrm{[\ion{Na}{I}/Fe]} & 6.24 &   0.30$\pm$0.24 \ \ (2) &  -0.13$\pm$0.22 \ \ (2) &  -0.30$\pm$0.20 \ \ (2) &  -0.73$\pm$0.29 \ \ (2) &  -0.63$\pm$0.22 \ \ (2) \\
\textrm{[\ion{Mg}{I}/Fe]} & 7.60 &   0.87$\pm$0.16 \ \ (3) &   0.02$\pm$0.17 \ \ (3) &   0.19$\pm$0.14 \ \ (3) &   0.44$\pm$0.23 \ \ (1) &   0.69$\pm$0.23 \ \ (1) \\
\textrm{[\ion{K}{I}/Fe]}  & 5.03 &   0.65$\pm$0.23 \ \ (1) &  <0.28$\pm$0.23 \ \ (1) &   0.55$\pm$0.24 \ \ (1) &   0.26$\pm$0.23 \ \ (1) &   0.64$\pm$0.24 \ \ (1) \\
\textrm{[\ion{Ca}{I}/Fe]} & 6.34 &   0.74$\pm$0.10    (13) &  -0.21$\pm$0.12 \ \ (5) &   0.09$\pm$0.14    (11) &   0.32$\pm$0.16 \ \ (6) &   0.22$\pm$0.11    (12) \\
\textrm{[\ion{Sc}{I}/Fe]*}& 3.13 &  -0.10$\pm$0.24 \ \ (1) &   0.00$\pm$0.22 \ \ (1) &   0.70$\pm$0.24 \ \ (1) &   0.00$\pm$0.24 \ \ (1) &   0.10$\pm$0.24 \ \ (1) \\
\textrm{[\ion{Ti}{I}/Fe]} & 4.95 &   0.22$\pm$0.21 \ \ (7) &   0.25$\pm$0.16 \ \ (5) &   0.42$\pm$0.24 \ \ (2) &           —    \ \     (0) &   0.40$\pm$0.21 \ \ (4) \\
\textrm{[\ion{Ti}{II}/Fe]}& 4.95 &   0.38$\pm$0.09 \ \ (4) &               — \ \ (0) &   0.90$\pm$0.12 \ \ (3) &           —    \ \     (0) &   0.20$\pm$0.15 \ \ (2) \\
\textrm{[\ion{Cr}{I}/Fe]} & 5.64 &  -0.18$\pm$0.21 \ \ (3) &   0.01$\pm$0.18 \ \ (3) &  -0.56$\pm$0.21 \ \ (2) &   0.03$\pm$0.27 \ \ (2) &  -0.58$\pm$0.21 \ \ (2) \\
\textrm{[\ion{Ni}{I}/Fe]} & 6.22 &  -0.13$\pm$0.20 \ \ (2) &   0.53$\pm$0.16 \ \ (5) &   0.07$\pm$0.14 \ \ (4) &  -0.01$\pm$0.27 \ \ (2) &  -0.12$\pm$0.16 \ \ (5) \\
\textrm{[\ion{Ba}{II}/Fe]}& 2.18 &  -1.10$\pm$0.15 \ \ (2) &  -0.95$\pm$0.18 \ \ (2) &  -0.39$\pm$0.12 \ \ (3) &  -0.43$\pm$0.17 \ \ (3) &  -1.20$\pm$0.16 \ \ (2) \\
\\
\textbf{NLTE }               \\
\textrm{[Fe/H]}           & 7.50 &  -2.69$\pm$0.15    (77) &  -2.02$\pm$0.12    (46) &  -2.37$\pm$0.12    (44) &  -2.27$\pm$0.20    (45) &  -2.35$\pm$0.14    (61) \\
\textrm{[\ion{Fe}{I}/H]}  & 7.50 &  -2.68$\pm$0.15    (73) &  -1.99$\pm$0.12    (41) &  -2.39$\pm$0.12    (42) &  -2.26$\pm$0.20    (43) &  -2.34$\pm$0.14    (55) \\
\textrm{[\ion{Fe}{II}/H]} & 7.50 &  -2.88$\pm$0.11 \ \ (4) &  -2.28$\pm$0.11 \ \ (5) &  -2.04$\pm$0.15 \ \ (2) &  -2.38$\pm$0.17 \ \ (2) &  -2.40$\pm$0.10 \ \ (6) \\
\textrm{[\ion{Na}{I}/Fe]} & 6.24 &  -0.17$\pm$0.24 \ \ (2) &  -0.75$\pm$0.22 \ \ (2) &  -0.53$\pm$0.20 \ \ (2) &  -1.12$\pm$0.29 \ \ (2) &  -0.97$\pm$0.22 \ \ (2) \\
\textrm{[\ion{Mg}{I}/Fe]} & 7.60 &   0.76$\pm$0.16 \ \ (3) &  -0.07$\pm$0.17 \ \ (3) &   0.12$\pm$0.14 \ \ (3) &   0.35$\pm$0.23 \ \ (1) &   0.61$\pm$0.23 \ \ (1) \\
\textrm{[\ion{K}{I}/Fe]}  & 5.03 &   0.19$\pm$0.23 \ \ (1) & <-0.20$\pm$0.23 \ \ (1) &   0.23$\pm$0.24 \ \ (1) &   -0.1$\pm$0.23 \ \ (1) &   0.29$\pm$0.24 \ \ (1) \\
\textrm{[\ion{Ca}{I}/Fe]} & 6.34 &   0.74$\pm$0.10    (13) &  -0.21$\pm$0.12 \ \ (5) &   0.10$\pm$0.14    (11) &   0.37$\pm$0.16 \ \ (6) &   0.24$\pm$0.11    (12) \\
\textrm{[\ion{Sc}{I}/Fe]*}& 3.13 &  -0.30$\pm$0.20 \ \ (1) &  -0.20$\pm$0.20 \ \ (1) &   0.50$\pm$0.20 \ \ (1) &  -0.20$\pm$0.20 \ \ (1) &  -0.10$\pm$0.20 \ \ (1) \\
\textrm{[\ion{Ti}{I}/Fe]} & 4.95 &   0.50$\pm$0.21 \ \ (7) &   0.57$\pm$0.17 \ \ (5) &   0.77$\pm$0.25 \ \ (2) &          —                \ \ (0) &   0.66$\pm$0.21 \ \ (4) \\
\textrm{[\ion{Ti}{II}/Fe]}& 4.95 &   0.27$\pm$0.09 \ \ (4) &               — \ \ (0) &   0.73$\pm$0.12 \ \ (3) &         —      \ \ (0) &   0.13$\pm$0.15 \ \ (2) \\
\textrm{[\ion{Cr}{I}/Fe]} & 5.64 &   0.08$\pm$0.21 \ \ (3) &   0.23$\pm$0.18 \ \ (3) &  -0.35$\pm$0.21 \ \ (2) &   0.29$\pm$0.27 \ \ (2) &  -0.33$\pm$0.21 \ \ (2) \\
\textrm{[\ion{Ni}{I}/Fe]} & 6.22 &  -0.32$\pm$0.20 \ \ (2) &   0.38$\pm$0.16 \ \ (5) &  -0.05$\pm$0.14 \ \ (4) &  -0.16$\pm$0.27 \ \ (2) &  -0.26$\pm$0.16 \ \ (5) \\
\textrm{[\ion{Ba}{II}/Fe]}& 2.18 &  -1.20$\pm$0.15 \ \ (2) &  -1.06$\pm$0.18 \ \ (2) &  -0.41$\pm$0.12 \ \ (3) &  -0.53$\pm$0.17 \ \ (3) &  -1.25$\pm$0.16 \ \ (2) \\
\bottomrule
\end{tabular}
\label{tab:figureabuns}
\end{table*}

\subsection{NLTE corrections}
\label{NLTE}

Departures from Local Thermodynamic Equilibrium (LTE) due to the radiation field in metal-poor red giants are known to impact the statistical equilibrium solution for elements like \ion{Fe}{I}.  These non-LTE (NLTE) effects can be large, significantly affecting stellar abundance solutions (e.g., $\Delta$log(X/H)$>0.2$).  To investigate the impact of NLTE corrections on our iron abundances, and other elements, we examine three databases and a paper; the MPIA\footnotemark{}\footnotetext{MPIA NLTE corrections: \url{http://nlte.mpia.de.}}  for \ion{Fe}{I} and \ion{Fe}{II} \citep{Bergemann2012}, \ion{Mg}{I} \citep{Bergemann2017}, \ion{Ca}{I} \citep{Mashonkina17}, \ion{Ti}{I} and \ion{Ti}{II} \citep{Bergemann2011}, and \ion{Cr}{I} \citep{Bergemann2010b}.  
We also examine the NLTE corrections available from the  INSPECT\footnotemark{}\footnotetext{INSPECT NLTE corrections: \url{http://inspect-stars.com.}} database for these corrections as listed in Table~\ref{tab:nltestub}, and INSPECT also includes NLTE corrections for \ion{Na}{I} \citep{Lind2012}. 
A comparison of the MPIA and INSPECT corrections for lines and elements in common suggests that the MPIA corrections are generally larger (more positive).
The NLTE corrections per line are included in Table~\ref{tab:nltestub}.

Finally, NLTE corrections for lines of \ion{Ba}{II} are from calculations by 
\citet{Mashonkina2019}, also available online\footnotemark{}\footnotetext{Mashonkina Ba II NLTE corrections: \url{http://www.inasan.ru/~lima/pristine.}}, and NLTE corrections for \ion{K}{I} are taken from \citet{Andrievsky10}.  
These are included in Table~\ref{tab:nltestub}. (The full table of NLTE corrections is currently in the Appendix).

\begin{figure*}
\begin{center}
    \setlength{\unitlength}{1cm}
    \includegraphics[width=17.5cm]{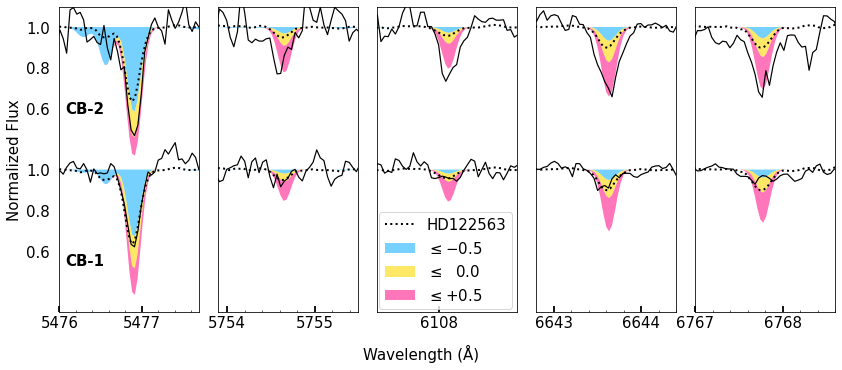}
    \caption{\ion{Ni}{I} lines for the two stars in Com\,Ber  compared to synthetic spectra with corresponding stellar parameters, which are rich in Ni (pink) and deficient in Ni (blue), to clarify the Ni-enrichment in the star CB-2.  }
    \label{fig:ni}
\end{center}
\end{figure*}

\section{Chemical Abundances}
\label{sec:chem}

The chemical abundances for five stars in three UFDs were uniquely determined from both a spectrum synthesis (Section~\ref{syntheses}) and an EW analysis (Section~\ref{EW}).  All abundances are reported in Table \ref{tab:figureabuns} and shown in Figs.~\ref{fig:full} and \ref{fig:bafe} compared to stars in the MW halo \citep{Aoki13, Yong13, Yong2021, Roederer14, Kielty2021}.
We also compare to stars in the UFDs Segue I \citep[][]{Frebel14} and Hercules \citep[][]{Koch08, Koch13, Francois16} because these UFDs have stars with similar metallicities and chemical abundances to Com\,Ber. 
The results from our EW and spectrum synthesis methods are in good agreement; however, our results suggest that the EW abundances have slightly higher precision.  For this reason, we apply the NLTE corrections to the EW-based abundances, and adopt those throughout the rest of this paper.


\subsection{Iron-Peak Elements}


The iron-peak elements in very metal-poor stars are expected to have formed primarily in core collapse supernova \citep{Tolstoy2009, HegerWoosley2010}.  At higher metallicities, iron-peak elements are also formed in SN Ia, either through the single or double degenerate scenarios \citep{Nomoto13}.  In this paper, we examine iron, chromium, and nickel.  

Iron is determined from $40-70$ \ion{Fe}{I} lines and $2-6$ \ion{Fe}{II} lines, where [Fe/H] is taken as the weighted average.   We find the abundances of \ion{Fe}{I} and \ion{Fe}{II} are in excellent agreement ($\le1\sigma$) for all stars except UMa\,I-1.
This star has a very low predicted gravity, \logg=0.3, which is slightly beyond the model atmosphere grid parameters. As such, we rounded \logg \, up to 0.5 for our analysis. This small offset may affect our \ion{Fe}{II} values slightly, but more importantly, we have poor constraints on \ion{Fe}{II} from only 2 lines.
%
%
Furthermore, the typical NLTE corrections for \ion{Fe}{I} are $\sim+0.1-0.2$, and $<0.1$ for \ion{Fe}{II}, which also provides a small improvement in the iron ionization equilibrium for UMa\,I-1.

Chromium abundances are determined from $2-3$ \ion{Cr}{I} lines, with significant NLTE corrections that raise these values.
We note that most analyses in the literature do not apply NLTE corrections, thus in Fig.~\ref{fig:full} we show both LTE and NLTE results for comparison.

Nickel is from $2-5$ \ion{Ni}{I} lines, where none have published NLTE corrections.  We find a surprisingly high [Ni/Fe] abundance in one star, CB-2.  This can be seen directly in the spectra as well (see Fig.~\ref{fig:ni}, and discussed further in Section~\ref{sec:chem}).

\subsection{Alpha Elements}

Alpha elements are even-Z elements that form primarily from helium nuclei captures during the carbon-, neon- and silicon-burning phases of massive star evolution, and through the $\alpha$-rich freeze-out during core collapse supernovae.  Recently, it has also been recognized that the interstellar medium may be enriched in $\alpha$-elements at early times through the winds of rapidly rotating massive stars
\citep[e.g.,][]{Limongi2018, Kobayashi2020}.
In this paper, the $\alpha$-elements are limited to magnesium, calcium, and titanium due to the low metallicity of our targets and the limited wavelength coverage of the GRACES spectrograph.

Magnesium abundances are determined from one weak line of \ion{Mg}{I} near 5528 $\Angstrom$, and the stronger \ion{Mg}{I}~b lines at 5172 and 5183 $\Angstrom$ (see Table~\ref{tab:linelist}).  The stronger lines are particularly sensitive to small uncertainties in the model atmospheres.
The NLTE corrections for all \ion{Mg}{I} lines in this analysis are small ($\sim-0.1$).

Calcium abundances are from $5-13$ \ion{Ca}{I} lines. The \ion{Ca}{II} triplet lines near $\sim$8000$\Angstrom$ were excluded because of their large EW and sizable NLTE corrections, despite being in a higher SNR wavelength region of our GRACES spectra. 
NLTE corrections are small, and typically similar to the average of the \ion{Fe}{I} lines, such that [Ca/Fe]$_{\rm NLTE} \sim$ [Ca/Fe]$_{\rm LTE}$.

Titanium is determined from up to 7 (4) \ion{Ti}{I} (\ion{Ti}{II}) lines, although no Ti lines were measurable in UMa\,I-2.  Both species of Ti have significant NLTE corrections, although in opposite directions; this can be seen in Fig.~\ref{fig:full} when both species are available.

\subsection{Odd-Z Elements}

Odd-Z elements are important indicators of core collapse supernova yields, as the difference in the energetic requirements for $\alpha$ particle capture versus neutron capture produces a noticeable odd-even effect in the predicted yields \citep{HegerWoosley2010, Takahashi18}. From our Gemini/GRACES spectra, we are able to observe a small handful of odd-Z element spectral lines, including sodium, potassium, and scandium (Na, K, and Sc).

%

Sodium abundances are from the two strong \ion{Na}{I}~D resonance lines near 5890 and 5895 $\Angstrom$.  These are present in all five of our target stars, where the stellar components are easily separated from the interstellar components, due to the radial velocities of our targets.  Not only are these lines strong, and therefore sensitive to small uncertainties in the model atmospheres, but they also suffer from significant NLTE effects as resonance lines.  We find NLTE corrections up to $\Delta_{\rm NLTE} \sim -0.5$ dex (see Table~\ref{tab:nltestub}).   We retain these lines for our analysis as the only measureable lines of an odd-Z element (Na), and note their sensitivities to the model atmosphere parameters represented by their larger errorbars.

Potassium is measured from the \ion{K}{I} 7699 \AA\ resonance line.  This region is significantly affected by telluric features (see Fig.~\ref{fig:spec}); however, the line is clear and present in most of our targets and comparison star HD~122563.  NLTE corrections are significant, $\sim -0.4$ dex, caused by an over-recombination on the first energy level of this spectral line in the atmospheres of late-type stars \citep{Andrievsky10}.


%
%

Scandium is measured from the \ion{Sc}{I} 5527 \AA\ line in all of our targets, as clearly seen in Fig.~\ref{fig:spec}. As an odd-Z element, it has significant hyperfine structure corrections, which are taken into account through the spectrum syntheses.
NLTE corrections are small, $\sim -0.2$ dex, taken from
\citet{Battistini15} per star.


\subsection{Neutron-capture Elements}

Neutron-capture elements are primarily formed through rapid-neutron capture in core collapse supernovae and neutron-star mergers, and slow-neutron capture during  thermal-pulsing in AGB stars.  The ratio of [Ba/Fe] produced in core-collapse supernovae (CCSNe) is expected to provide an r-process Ba floor, with contributions from the other processes building above that.  To ascertain these different nucleosynthetic sources, often it is necessary to observe a pure r-process element (usually Eu).
Unfortunately, Eu has only a very weak line near 6645 $\Angstrom$ in the GRACES spectra, which is too weak to measure at our SNR.  The stronger \ion{Eu}{II} line near 4129 $\Angstrom$ is outside the wavelength range of GRACES. 

Barium is primarily determined from two \ion{Ba}{II} lines near 6141 and 6496 $\Angstrom$, and occasionally the weaker \ion{Ba}{II} line near 5853 $\Angstrom$.  This element requires corrections for hyperfine structure and isotopic splitting, taken into account in our analysis through spectrum syntheses.  
NLTE corrections are from \citet{Mashonkina2019Ba}.




\begin{figure*}
\begin{center}
    \setlength{\unitlength}{1cm}
    \includegraphics[width=17.5cm]{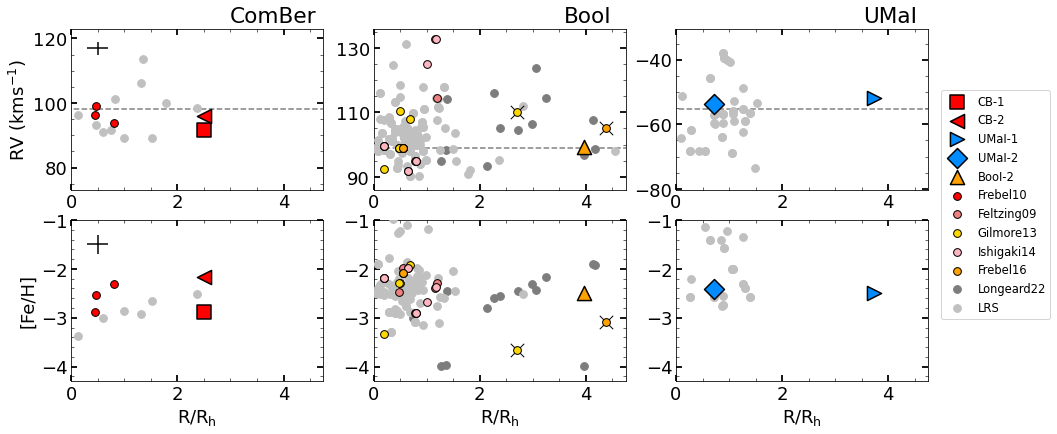}
    \caption{ Radial distance in units of the half-light radius for all apparent members wtih spectroscopic studies (see online table for full list). Coloured points with black outlines are from HRS analyses. Low- and medium-res observations (LRS) are grey points. Dashed line shows the mean heliocentric radial velocity of the galaxy (see Table~\ref{tab:galpar}). Stars with HRS beyond $R\gtrsim2 R_h$,
    are additionally marked with a black ``X" (they are members). Different observations of the the same star are {\it not} plotted twice,  only the most recent data is shown. Sample uncertainty in $R$ is from the uncertainty in galactic half-light radius from \citet{McConnachie2012}. Sample uncertainties of $\Delta$[Fe/H] = 0.2 and $\Delta$RV = 2 \kms are shown in the left panels only.}
    \label{fig:RVFe}
\end{center}
\end{figure*}

\begin{figure*}
	\includegraphics[width=17.5cm]{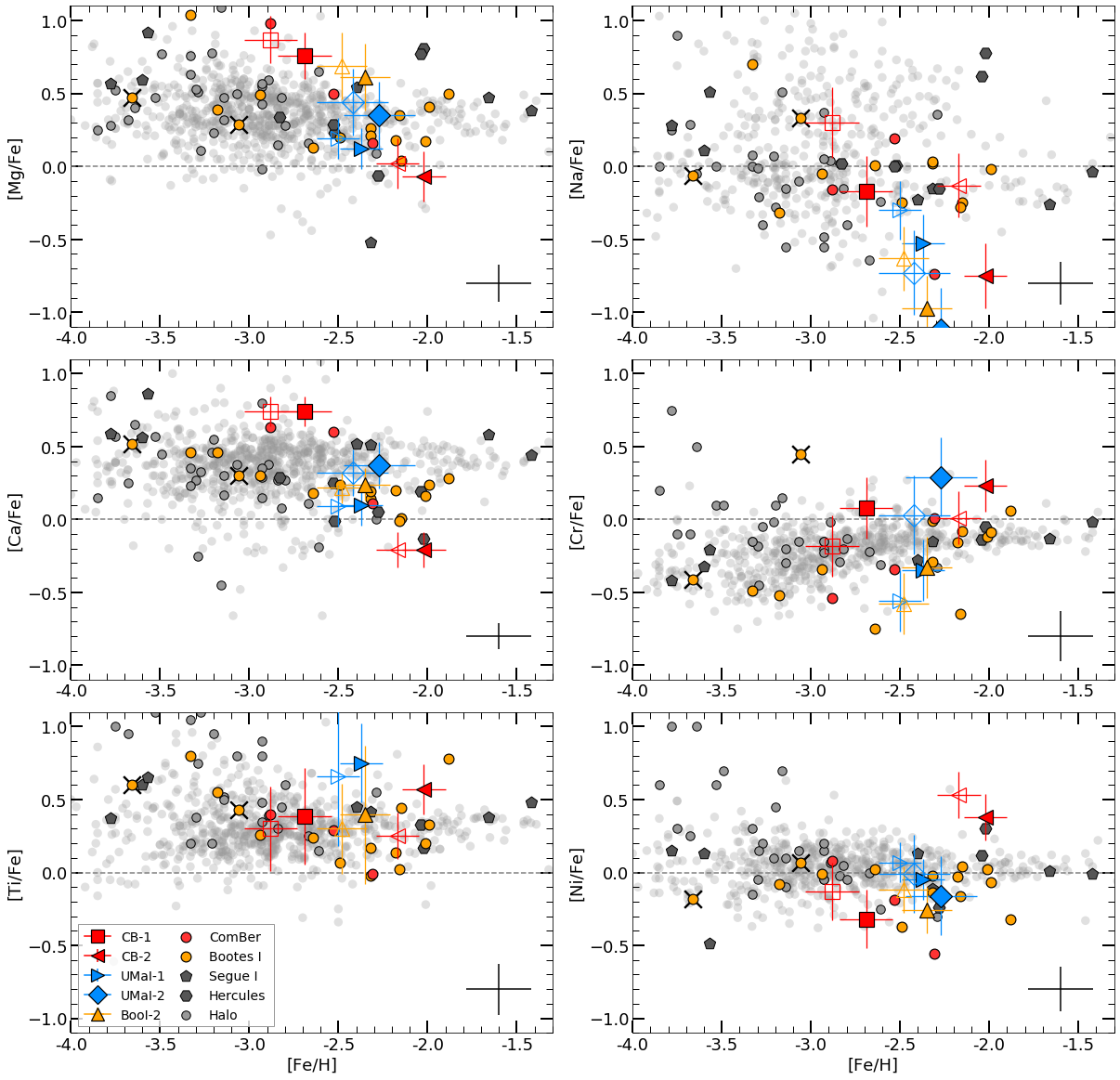}
    \caption{Markers with dark outline indicate hi-res observations, except for halo stars, for which it indicates the same analysis tools used as in this paper. 
    Stars with model atmospheres analyses at $\gtrsim3$ R$_{\rm h}$, Boo-980 ([Fe/H]$=-3.1$), and Boo-1137 ([Fe/H]$=-3.7$), are additionally marked with a black x. 
    For our data, filled: NLTE, and unfilled: LTE. Most reference data is LTE. Ti values were averaged over \ion{Ti}{I} and \ion{Ti}{II}, and large differences between the two are shown in the large error bars.
    $\alpha$-element abundances drop to solar at [Fe/H]~-2.2, with new observations in Mg and Ca consistent. 
    (Halo: \citealt{Yong13, Yong2021, Aoki13, Roederer14}, Halo marked with black outlines \citealt{ Kielty2021}; Com\,Ber: \citealt{Frebel10}; Boo\,I: \citealt{Norris10c, Ishigaki14, Gilmore13, Frebel16} ; Hercules: \citealt{Koch08, Koch13, Francois16}; Segue\,I: \citealt{Frebel14}.
    \label{fig:full}}
\end{figure*}

\section{Chemical History of Ultra-Faint Dwarfs}
\label{sec:evol}

\subsection{Membership in the UFD hosts}

In this paper, we present high resolution spectroscopic analyses for five stars in three ultra faint dwarf galaxies; Com\,Ber, UMa\,I, and Boo\,I.  
This project was partially motivated as a test of our target selection methodology, described in \citealt{McVenn2020}, and its application to targets well beyond the tidal radius of their host galaxy.  
The radial velocities and low metallicities of our targets suggest that each is associated with their host UFD galaxy, providing excellent validation of our Bayesian inference method for
finding highly probable members in UFDs from Gaia photometry and astrometry, with or without {\sl a priori} radial velocities or metallicity information.

In Fig.~\ref{fig:RVFe}, the radial velocities (RVs) and metallicities ([Fe/H]) of our targets are shown relative to others in the literature as a function of half-light radius in each UFD.  Our metallicities are consistent with stars located in the central regions.  A similar result was seen in Boo\,I by \citet{Longeard22}, who selected targets\footnote{We note that nearly all of the targets selected by \citet{Longeard22} have P$_{\rm sat}>0.75$ when using our selection criteria, which did not include metallicity. Two of their other outermost stars have P$_{\rm sat}\sim0.6$, and only their most distant and metal-poor star ([Fe/H]$\sim-4$ at $\sim4$\,R$_{\rm h}$) had P$_{\rm sat}\sim0.2$.} 
for MRS from the Pristine survey \citep{Starkenburg2017b}. 
The positions of the members in Fig.~\ref{fig:halflight} do not suggest that Com\,Ber or Boo\,I are surrounded by elongated streams; however, the  distribution of stars along the major-axis of UMa\,I is quite interesting.  This was also noticed by \citet{Simon2019}, and is strengthened by our analysis of the outermost member (UMa\,I-1).

For comparison,
\citet{Chiti21} interpret the presence of one member of the UFD Tuc\,II at $\sim$9 R$_h$ as evidence for either tidal stripping through interactions with the MW halo, strong bursty feedback during star formation that can provide a kick to the stellar dynamics, or possibly the remnant of an early dwarf galaxy minor merger.  
%
%
%
%
%
We suggest that deeper searches into the outskirts of the UFDs, now possible with the exquisite Gaia DR3 data, may provide new opportunities to study minor mergers in dwarf galaxies to recover the conditions in very low mass and pristine systems from the earliest epochs.


\begin{figure*}
	\includegraphics[width=17cm]{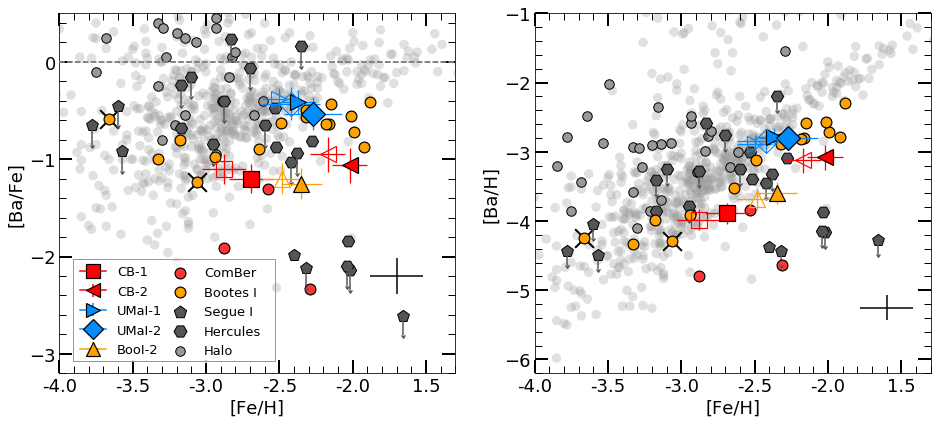}
    \caption{ [Ba/H] shows flat r-process floor at -4 for system unenriched by s-process, which is mostly seen in Segue I.  Symbols are the same as in Fig.~\ref{fig:full}. Ba II in Com\,Ber follows low abundances consistent with r-process ratio, showing evidence of no AGB evolution enrichment, and enrichment consistent with one SNII enrichment event. Symbols are the same as in Fig.~\ref{fig:full}.
    \label{fig:bafe}}
\end{figure*}

\subsection{Chemical Abundances in the UFDs}

High-resolution hydrodynamic simulations of UFD galaxies suggest that the baryonic component of these dark matter dominated systems are characterized by bursty star formation histories, truncated early
by reionization.  The stars formed during these bursts will be strongly clustered in location and chemically similar  \citep{Wheeler2019, Revaz2018, Applebaum2021}.  
%
%
Deep color-magnitude diagrams provide observations that identify the stellar populations observed in the UFDs as old ($>$10 Gyr) and very metal-poor ([Fe/H]$<-2$; \citealt{McConnachie2012, Weisz2014, Brown2014, Laevens2015}). 
Thus, UFDs have been proposed as fossils from the epoch of reionization, having undergone very little evolution since that time.

The simulations further predict that the stars we observe today (z=0) in the UFDs may include the chemical signatures of  Population III stars and the earliest supernovae \citep{Tumlinson2010, Frebel2012, Hartwig18, Ishigaki2018}.
It may be possible to use detailed abundance patterns to constrain the mass of the first stars to pollute these systems.  For example, very massive pair-instability supernovae (PISNe) predict stars with unusually deficient neutron-capture element yields (e.g., Ba) and high ratios of [Ca/Mg] \citep{Takahashi18}, along with the classic odd-even effect \citep{HegerWoosley2010}.  Alternatively, lower mass metal-poor events can produce "faint supernova", which underproduce both the iron-peak and heavier elements, as well as the brightness of the supernova event, as $^{56}$Ni falls into the core during collapse \citep{Nomoto13}.

One challenge to UFD simulations has been the predicted metallicities of the lowest mass systems, and thereby the observed metallicity distributions, let alone the detailed chemical abundances.  For example, \citet{Wheeler2019} were able to reproduce many observations of the lowest mass dwarf galaxies (mass, M/L ratio, size, frequency, star formation histories), however they were not able to reproduce the observed stellar metallicities.  Simulated metallicities were too low, as gas is lost from low mass systems through supernova feedback.  Multiple suggestions were offered, including pre-enrichment from Population III stars, pre-enrichment from a more massive host galaxy, and varying nucleosynthetic yields \citep{Revaz2018, Wheeler2019}, or that the feedback adopted was too efficient \citep{Agertz2020}.  
More recent zoom-in cosmological simulations of the UFDs by \citet{Applebaum2021} find better agreement between stellar metallicities and simulations when total metals (Z) is used, which includes all elements (dominated by oxygen rather than iron).  Thus, a metallicity floor of log (Z/Z$_\odot$) $=-4$ is more successful at reproducing the observed metallicities in the UFDs than a floor of [Fe/H]$=-4$.

Some UFD simulations have sufficient numbers of stars to show a metallicity distribution. 
Nucleosynthetic events will enrich a local bubble of the interstellar gas, resulting in significant dispersions in the star-to-star metallicities and [X/Fe] ratios from stochastic sampling of the interstellar gas \citep[e.g.,][]{Applebaum2021, Kobayashi2020, Revaz2018, Fenner2006}.  
However, metallicity distributions or {\sl gradients} may also signal evidence for a dwarf galaxy halo, formed through minor merging of sub-halos \citep{BenitezL2016, Deason2022}.
In the rest of this section, we discuss the chemical abundances of the three UFDs in this paper in terms of their physical properties and evolutionary histories.

\subsubsection{Com Ber}

Com\,Ber is a UFD with total mass $M$ $\approx$ 1.2 $\times$ $10^6$ \msun\ \citep{Simon2007}, apparent and absolute magnitudes of $m_V$ $\approx$ 14.1 and $M_V$ $\approx$ -4.1, half-light radius $R_{\rm h}$ $\approx$ 65 pc, and at a distance of $D$ $\approx$ 40 kpc \citep{McConnachie2012}.
\citet{Simon18}  suggests, based on Gaia DR2, that Com\,Ber is currently within 4 kpc of its pericenter.
A study of the 
star-forming history of Com\,Ber using HST photometry suggests that most (>75\%) of its stars may have formed before the epoch of reionization \citep{Brown2014}.

Three bright stars ($V$ $<18.1$) were analysed from high resolution spectroscopic data by \citet{Frebel10} who measured elemental abundances (C to Zn), and some neutron-capture elements (Sr to Eu). They found a significant metallicity dispersion from $-3 <$ [Fe/H] $<-2$, and an unusually low neutron-capture element ratios in two stars with [Fe/H]$\sim-2$.  It was unclear how such "high" iron abundances could have happened without contributions to the heavier elements.  Similar abundance patterns were seen in other UFD galaxies, leading \citet{Frebel2012} to
develop a set of stellar abundance signatures that could be the products of chemical "one-shot" events, where only one (long-lived) stellar generation forms after the first Population III SN explosions.  For Com\,Ber, the chemical abundances were used to propose that a single, metal-free $\sim20$ M$_\odot$ core collapse supernova  produced the abundance distribution observed.   
\citet{Sitnova21} conducted a NLTE re-analysis of the three stars from \citet{Frebel10} and find amongst the lowest [Na/Fe] abundances measured in any stars to date, which they interpret as an odd-even footprint of nucleosynthesis in Pop III stars.
We note, as seen in 
Fig.~\ref{fig:RVFe}, that these three stars are all centrally located ($<$1 R$_{h}$).

Subsequently, 
\citet{Vargas13}
reported a higher metallicity spread ($-3.4 <$ [Fe/H] $<-2$) and a spread in  $\alpha$-elements ($-0.9 <$ [$\alpha$/Fe] $< 0$) from 10 more stars from med-res spectroscopy. They conclude that the low [$\alpha$/Fe] ratios at the highest [Fe/H] is caused by iron contributions from SNIa.  This is difficult to reconcile with the "one-shot" model above.

In this analysis, we present HRS of two new stars on the outskirts of Com\,Ber ($\sim$2.5 R$_{\rm h}$). These two stars span the same metallicity range as the three previously studied inner stars by \citet{Frebel10}, i.e., $-3 <$ [Fe/H] $<-2$.  This strongly suggests that they formed at the same time as the inner stars, perhaps arriving at their current locations either through tidal stripping with the MW halo or bursty feedback during star formation \citep[i.e.,][]{Chiti21}.
To examine these options further, we evaluate the detailed chemical abundances from HRS studies of the Com\,Ber stars. Our work is presented as both LTE and NLTE abundances in Figs.~\ref{fig:full} and \ref{fig:bafe} for ease of comparison with the previous LTE analysis.  We notice three particular chemical signatures;

\begin{itemize}
    \item All five Com\,Ber stars follow a tightly decreasing [Mg/Fe] and [Ca/Fe] (and possibly [Na/Fe], acting as an $\alpha$-element), with increasing [Fe/H].  The relationship is so tight that it strongly favours late iron contributions from SN Ia as originally suggested by \citet{Vargas13}.   We also note that this is in contrast to the relationships between [Mg,Ca/Fe] for the stars in Segue\,I and possibly Hercules.
    \item The [Ni/Fe] abundances also seem to decrease with [Fe/H], similar to the Ni-Na signature seen in other dwarf galaxies and MW accreted stars \citep{Nissen1997, Nissen10, Venn04}. Only the most metal-enhanced star CB-2 deviates from this trend, where [Ni/Fe]$\sim+0.5$.  This is a very unusual signature, verified by the strength of the \ion{Ni}{I} lines in Fig.~\ref{fig:ni}. One possibility might be the late contributions from sub-Chandrasekhar mass (sub-$M_{\rm Ch}$) white dwarfs in double-degenerate SNIa \citep[e.g.,][]{Kobayashi2006,Kobayashi2020Ni}.
    \item The [Ba/Fe] abundances in our outermost stars are not as low as some of the previously analysed inner-most stars. Again, this is difficult to reconcile with the currently favoured "one-shot" chemical evolution model for Com\,Ber. This is in contrast to [Ba/Fe] in Segue\,I, and in better agreement with the majority of stars analysed in Hercules.
\end{itemize}

Given the detailed chemical signatures for 3 inner and 2 outer stars in Com\,Ber, we favour a scenario where these stars formed together in the inner regions with sufficient time for SN\,Ia contributions.  
This explains the position in [Fe/H] of the [$\alpha$/Fe] knee, known to be at lower metallicities in dwarf galaxies with slower star formation histories \citep{Tolstoy2009, Hasselquist2021}. 
%
This also suggests that the outermost stars were
relocated to their current outer positions
{\sl after} contributions in the central regions from SN\,Ia.  
The unusually low [Ba/H] ratios for two of the innermost stars remains a puzzle; one option may be due to inhomogeneous mixing of contributions from metal-poor AGB stars.  If the rise in [Ba/H] (seen in Fig.~\ref{fig:bafe}) is from slow-neutron captures in AGB stars, whose yields depend on both stellar mass and metallicity \citep{Herwig05, Nomoto13}, then a wide range in [Ba/H] is possible.
Combining the similar
timescales for AGB and SNIa contributions, this implies that the mixing timescales of these products into star forming gas could have been very stochastic to result in low [Ba/H] at higher [Fe/H] values. 
The Ni enrichment in our outermost and Fe-enriched star supports this suggestion further, as contributions from sub-Ch mass white dwarfs would also require additional time.
Thus, we suggest an alternative explanation to the ``one shot" model for the low [Ba/Fe] ratios seen at higher [Fe/H] values for two inner stars, i.e., stochastically sampled yields from metal-poor AGB stars and SNIa in an inefficient star forming environment (i.e., an UFD).
Again, we suggest the outermost stars moved to their current locations {\sl after} these phases of chemical evolution.

\subsubsection{Bo{\"o}tes I}



Boo\,I is a UFD with total mass $M$ $\approx$ 1.1 $\times$ $10^7$\msun\ \citep{Munoz06}, half-light radius $R_{\rm h}$ $\approx$ 217 pc, and is fairly close to the MW at a distance of $D$ $\approx$ 66 kpc \citep{McConnachie2012}. Boo\,I is one of the brightest UFDs, with $L\approx10^5L_{\odot}$ \citep{Munoz06}, with apparent and absolute magnitudes of $m_V$ $\approx$ 12.8 and $M_V$ $\approx$ -6.3 \citep{McConnachie2012}. Because of this, Boo\,I is one of the most observed UFDs with nearly 70 confirmed members, and high-resolution spectroscopic chemical abundances for 18 stars  \citep{Feltzing09, Norris10a, Gilmore13, Ishigaki14, Frebel16}; this can also be seen in Fig.~\ref{fig:halflight}.  

Photometric studies have found that Boo~I has a narrow CMD at the MSTO, indicating an old single-age population  \citep[13.7 Gyr;][]{Okamoto12}.
However, HST photometry suggests there may be two ancient populations with a very small age spread \citet{Brown2014}. 
A kinematic study by \citet{Koposov11} found two distinct (hot and cold) components with different velocity dispersions and mean metallicities.  
Furthermore, \citet{Belokurov06} and \citet{Okamoto12} noted an irregular morphology which may be related to tidal disruption. 
This was confirmed and studied further with confirmed outer members by \citet{Longeard22}.

To this discussion, we add HRS for one new outer star, at $\sim4$R$_{\rm h}$.  This is one of the most distant stars in Boo~I, and notably it is on the opposite side from the other two outermost stars with HRS.
Its radial velocity and low metallicity are consistent with the inner region stars.
Overall, the three outer stars with HRS have a metallicity distribution similar to the inner stars, although some stars in the central regions are slightly metal-enhanced
(see Fig.~\ref{fig:RVFe}).
The metal-enhanced inner stars may suggest that a small amount of ongoing star formation and chemical enrichment occurred at
later times.
An interesting alternative to this simple scenario is the possibility that the outer stars represent a metal-poor stellar halo, e.g., from the collapse of an inhomogeneous system of star forming clouds, or even a dwarf galaxy minor merger \citep[e.g.,][]{BenitezL2016, Deason2022}.
 
To explore this further, we examine the detailed chemical abundances of the stars associated with Boo\,I. 
\citet{Norris10a}, \citet{Lai11}, and \citet{Gilmore13} showed that the stars in Boo~I have a wide range in metallicity 
($-3.8 <$ [Fe/H] $< -1.8$) 
and a large spread in [C/Fe]
(although \citealt{Ishigaki14} showed this is restricted to the very metal-poor stars with [Fe/H] $< -2.7$). 
\citet{Gilmore13} suggested two discrete paths for chemical enrichment at low metallicity in Boo\,I; one where CEMP-no (carbon-enhanced metal poor stars that show no enrichment in s- or r-process) stars form rapidly and early, and a second path that forms carbon-normal stars.  They suggested these
two paths may be from different dwarf galaxy progenitors which merged, or may be due to inhomogenous mixing of the SN ejecta in the ISM during subsequent star formation events.


Looking into these two scenarios, we note that the outer stars studied by \citet{Lai11} are all carbon-normal, and the three outer stars with HRS in Boo\,I are
%
%
also carbon-normal. In fact, their full chemical analyses suggest they are similar to one another (in [X/Fe]), as well as to other very metal-poor stars in the MW halo.  This includes our star (BooI-2) as shown in Figs.~\ref{fig:full} and \ref{fig:bafe} (with the exception of Na), and we do not see any evidence for carbon enhancements, e.g., no strong C$_2$ Swan bands near 5200\AA\ nor CH/CN features near 4300 \AA.  Similarly, the most metal-poor star associated with Boo\,I (Boo-1137; \citealt{Norris10a}) has normal MW halo-like abundances extending from C through Zn, including the heavy elements Sr and Ba.
%
%
Finding that all of the outer stars analysed spectroscopy are carbon-normal is not necessarily inconsistent with the low percentage of CEMP stars in Boo\,I.  \citet{Lai11} suggested that only 12\% of the Boo\,I stars are CEMP, thus finding four outer stars that are metal-poor and C-normal is statistically consistent with being drawn from the inner sample, e.g., through tidal stripping.
Alternatively, as they also appear to be chemically consistent with one another, formation in a separate dwarf galaxy that merged later is also plausible
 \citep[e.g.,][]{BenitezL2016, Deason2022}.
%
An HRS analysis of more metal-poor stars in the outskirts of Boo\,I could provide indicators of
formation in a previously independent and lower mass dwarf galaxy (e.g., a lower [$\alpha$/Fe] knee). 
This could help to address whether this UFD underwent a minor merger that provided stars that now occupy a halo.

As a final comment, two additional chemical evolution models for Boo\,I suggest that very inefficient star formation turned less than 3\% of the baryons into stars \citep{Vincenzo14} and that bursty star formation episodes and supernova events were not the dominant sources of the gas removal \citep{Romano15}. The latter also suggested that
external mechanisms are needed to model this UFD (e.g., the absence of neutral gas,  \citealt{Bailin2007}), in addition to the outermost stars.

 



 \begin{figure}
	\includegraphics[width=8.5cm]{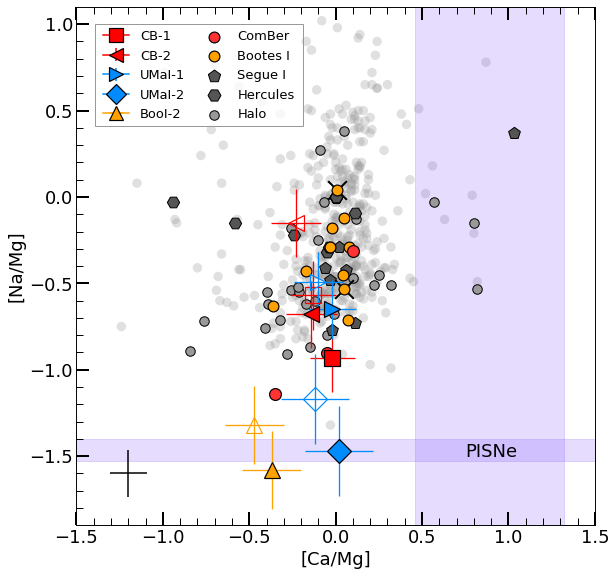}
    \caption{The [Ca/Mg] ratios as a function of [Na/Mg]. The shaded regions correspond to predicted PISNe yields \citep[][]{Takahashi18}. Two stars, UMa1-2 and BooI-2, have [Na/Mg] ratios consistent with PISNe predictions, but their [Ca/Mg] ratios are significantly lower than predictions. Thus, none of our stars nor Boo\,U stars from the literature match the PISNe signatures. 
    \label{fig:camg}}
\end{figure}

\subsubsection{UMa I}

UMa\,I is a UFD with total mass $M$ $\approx$ 1.5 $\times$ $10^7$\msun\ \citep{Simon2007}, apparent and absolute magnitudes of $m_V$  $\approx$ 14.4 and $M_V$  $\approx$ -5.5, half-light radius $R_{\rm h}$ $\approx$ 235 pc, and it is at a distance of $D$ $\approx$ 97 kpc \citep{McConnachie2012}.
HST photometry showed that UMa\,I has an age within 1 Gyr of the oldest known MW globular cluster, M92 \citep{Brown12}. 
Furthermore,  \citet{Brown2014} reconstructed its star formation history with a two-burst model, such that roughly half of the stars in UMa\,I were formed over 13 Gyr ago, and the other half by $z \sim 3$ (11.6 Gyr ago).
They also suggest that a physical distortion in the structure of UMa\,I may account for the apparent age spread, e.g.,
a tidal distortion, first noticed in a photometric study by \citet{Okamoto08}, 
A dynamical analysis using radial velocities and Gaia DR2 proper motion data found that UMa\,I is within 4 kpc of pericenter \citep{Simon18, Fritz18b},
and may even be a 
“backsplash” satellite, in that its orbit may have taken it beyond the virial radius of the MW dark matter halo.
For a galaxy at this distance, no more than
15-20\% of its stars are thought to be vulnerable to tidal stripping \citep{Simon2019}.
However, the elongated shape of UMa\,I is distinct and upon first glance many suspect this is a result of tidal interactions. As seen in \citet{Asya2022}, pericenter estimates can be very sensitive to measured distances and proper motions, therefore decisions based on the locations of the outer stars in UMa\,I as a result of tidal stripping should be reserved until precise orbits have been determined.

UMa\,I is slightly further than Com\,Ber or Boo\,I, so previous studies have been restricted to MRS analyses, or radial velocities only from HRS \citep{Kleyna05, Simon2007, Martin07, Vargas13}.  
Here, we provide the first HRS model atmospheres analysis of two stars in UMa\,I. Radial velocities are in good agreement with MRS estimates in the literature.
%
Our two UMa\,I targets include one inner star (< 1R$_{\rm h}$) and one in the outskirts ($\sim$ 4R$_{\rm h}$); see Fig.~\ref{fig:RVFe}. 
The two stars have similar metallicities and chemical abundances, suggesting that 
they formed from the same or similar interstellar gas.


Some of the inner stars in UMa\,I may be more metal-rich than the two in this analysis (see Fig.~\ref{fig:RVFe}), indicative of on-going or bursty star formation with inhomogeneous mixing of the gas in the central regions.  The position of our outermost star could then be explained by 
bursty feedback associated with this star formation epoch.
%
%
%
%
The very low [Na/Fe] abundances we find in our two UMa\,I stars also favour a later formation time, if the low values observed are due to later contributions to Fe-only from SN\,Ia (as discussed above for Com\,Ber). 
More members of UMa\,I with a wider range in metallicity will need to be analysed to comment further on the evolutionary history of UMa\,I.

\subsection{No PISN, No NSM}

If the UFDs are dominated by old stellar populations, i.e., fossil relics from before reionization, 
then simulations predict that the stars we observe today
($z$=0) may include traces of the chemical signatures of very massive Population III stars 
\citep{Tumlinson2010, Frebel2012, Hartwig18, Ishigaki2018}.
Pair-instability supernova (PISN) are a common fate for such massive stars, 
Predictions for nucleosynthesis that can take place during a PISN show unique patterns, e.g., high ratios in the certain light elements (e.g,. [Ca/Mg]), and the absence of heavy elements  (e.g., [Ba/Fe]), \citealt{Takahashi18}), along with the classic odd-even
effect seen in the [Na/Mg] or [Al/Mg] ratios \citep{HegerWoosley2010,Salvadori19}.
When the star formation history has been as slow and inefficient as in UFDs, it is 
reasonable to look for these chemical signatures since we expect very little later star formation to have affected the chemical evolution of the system.
%
%
%
In Fig.~\ref{fig:camg}, the high [Ca/Mg] ratios predicted by \citet{Takahashi18} are compared to the very low and narrow [Na/Mg] predictions.  We do not find any stars with both high [Ca/Mg] and low [Na/Mg], suggesting that the abundance patterns in these UFDs are not dominated by the ejecta from PISN.

We also mention that contributions from compact binary mergers are predicted to produce very high abundances of neutron capture elements \citep[e.g.,][]{Korobkin2012, Wu2016}, as seen in the UFD galaxy Ret~II \citep{Roederer16, Ji16, Ji2018}.  No stars in these three UFD galaxies appear to have high [Ba/Fe] ratios; thus, we find no evidence for such signature of compact binary mergers in these systems.


\section{Conclusions}

Ultra-faint dwarf galaxies are unique objects for studying galaxy formation and the early universe. This is because their old stellar populations and nearby locations in the MW halo mean that analyses of individual metal-poor stars may provide clues to physical conditions in the early Universe.
In this paper, we present high-resolution spectra taken with the Gemini/GRACES spectrograph of five stars associated with the Coma Berenices, Bo\"otes I, and Ursa Major I ultra faint dwarf galaxies. Most of these stars are in the outskirts, > 2R$_{\rm h}$, but include one inner star in UMa\,I.
Their radial velocities and metallicities are consistent with membership in these UFDs, showing that the Bayesian inference method using Gaia photometric and astrometric data  \citep{McVenn2020} is highly successful at selecting members, even at large radial distances.
 
The detailed model atmospheres analyses of these spectra have provided additional information on the evolution of these UFDs.

\begin{itemize}
\item The [$\alpha$/Fe] abundances in Com\,Ber suggest this UFD has been enriched in iron from SN\,Ia. This is at odds with the previous suggestion that Com\,Ber is a fossil relic enriched by only one early core-collapse SN (i.e., the ``one-shot" model).  A star formation model that includes contributions from SN\,Ia is supported by an elevated Ni in the least metal-poor star, possibly due to enrichment from sub-$M_{\rm Ch}$ mass white dwarfs.  The [Ba/H] ratios
vary from star to star in Com\,Ber, which may be more consistent with inhomogeneous mixing of AGB yields in the gas during star formation.  Finally, these conclusions include the outermost stars, which suggests those stars moved to their outer locations after formation in the central regions.

\item
In Bo\"otes I, the analysis of our star and two others with HRS at > 3R$_{\rm h}$ show they are metal-poor and C-normal. Some inner stars may be slightly more metal-rich implying longer-lived star formation in the centre; alternatively, several newly discovered very metal-poor stars in the outerskirts suggests the halo remnant of a dwarf galaxy minor merger.

\item
We present the first model atmospheres abundance analysis of two stars associated with UMa\,I; one inner star ($<$1 R$_{\rm h}$) and one in its outskirts ($R$ $\sim$ 4R$_{\rm h}$). The outermost star is aligned with the distorted shape of UMa\,I seen in its photometry.  Their metallicities and chemical signatures are similar to one another, as well as the stars in other UFDs. Very low [Na/Fe] values are seen, similar to the stars in Com Ber, which may indicate a star formation history that permits contributions from SN\,Ia events.  This suggests the outermost star moved to its current locations after its formation with the inner star in the central regions.
\end{itemize}


While single star observations are time consuming and expensive for major observatories, spectroscopic observations of carefully selected stars in the outskirts of the UFD galaxies are necessary to address questions related to the formation of the MW, dwarf-dwarf galaxy mergers, the search for the first stars, the chemical yields from metal-poor supernovae, and the chemical evolution of these unique systems.  We look forward to the new opportunities provided by the forthcoming Gemini High-Resolution Optical SpecTrograph (GHOST, \citealt{Pazder2020}).

\section*{Acknowledgements}

We acknowledge and respect the l\textschwa\textvbaraccent {k}$^{\rm w}$\textschwa\ng{}\textschwa n peoples on whose traditional territory the University of Victoria stands and the Songhees, Esquimalt and $\ubar{\rm W}$S\'ANE\'C peoples whose historical relationships with the land continue to this day.

The authors wish to recognize and acknowledge the very significant cultural role and reverence that the summit of Maunakea has always had within the indigenous Hawaiian community.  We are very fortunate to have had the opportunity to conduct observations from this mountain.

This work is based on observations obtained with Gemini Remote Access to CFHT ESPaDOnS Spectrograph (GRACES), as part of the Gemini Large and Long Program, GN-2020B-LP-102.
Gemini Observatory is operated by the Association of Universities for Research in Astronomy, Inc., under a cooperative agreement with the NSF on behalf of the Gemini partnership: the National Science Foundation (United States), the National Research Council (Canada), CONICYT (Chile), Ministerio de Ciencia, Tecnolog\'{i}a e Innovaci\'{o}n Productiva (Argentina), Minist\'{e}rio da Ci\^{e}ncia, Tecnologia e Inova\c{c}\~{a}o (Brazil), and Korea Astronomy and Space Science Institute (Republic of Korea).
CFHT is operated by the National Research Council of Canada, the Institut National des Sciences de l'Univers of the Centre National de la Recherche Scientique of France, and the University of Hawai'i. ESPaDOnS is a collaborative project funded by France (CNRS, MENESR, OMP, LATT), Canada (NSERC), CFHT, and the European Space Agency. 
Data was reduced using the CFHT developed OPERA data reduction pipeline.

This work has also made use of data from the European Space Agency mission Gaia (https://www.cosmos.esa.int/gaia), processed by the Gaia Data Processing and Analysis Consortium (DPAC, \url{https://www.cosmos.esa.int/web/gaia/dpac/consortium}). Funding for the DPAC has been provided by national institutions, in particular the institutions participating in the Gaia Multilateral Agreement. This research has made use of use of the SIMBAD database, operated at CDS, Strasbourg, France (Wenger et al. 2000).

The authors would like to thank Asya Borukhovetskaya in particular for her consultation and on dynamics and tidal effects which were invaluable for our understanding of the galaxies discussed here. 



\section*{Data Availability}
GRACES spectra are available at the Gemini Archive web page \url{https://archive.gemini.edu/searchform}. The data underlying this article are available in the article and in its online supplementary material. 

\bibliographystyle{mnras}
\bibliography{bibliography.bib}


\newpage

\section{Tables Appendix}

\tabcolsep=0.16cm
\begin{table}
\caption{$\delta T$,   $\delta g$,   $\delta m$, and $\delta \xi$ are systematic errors in the stellar parameters, \teff, \logg, metallicity and microturbulence. $\sigma_{EW}$ and $\sigma_{syn}$ denotes the line scatter for the equivalent widths and syntheses methods respectively. For species with fewer than 5 lines, the scatter from \ion{Fe}{I} was adopted, and was reduced by $\sqrt{n}$ of the species. $\Delta$EW and  $\Delta$syn are the linear combinations of the line scatter and the systematic errors. }
\begin{tabular}{lrrrrrrrr}
\toprule
\textbf{species} & $\Delta T$ &   $\Delta g$ &   $\Delta m$ &   $\Delta \xi$ &  $\sigma_{EW}$ & $\sigma_{syn}$ &  $\Delta$EW &  $\Delta$syn \\
\midrule
\textbf{CB-1}    &       &       &       &       &        &           &             &       \\
\textbf{Fe I   } &  0.14 &  0.00 &  0.04 &  0.01 &   0.02 &      0.05 &        0.15 &       0.15 \\
\textbf{Fe II  } &  0.04 &  0.02 &  0.01 &  0.01 &   0.10 &      0.14 &        0.11 &       0.15 \\
\textbf{Na I   } &  0.18 &  0.01 &  0.06 &  0.03 &   0.14 &      0.17 &        0.24 &       0.26 \\
\textbf{Mg I   } &  0.10 &  0.04 &  0.03 &  0.02 &   0.11 &      0.13 &        0.16 &       0.17 \\
\textbf{\ion{K}{I}}&0.11 &  0.01 &  0.03 &  0.01 &   0.20 &      —    &        0.23 &       —    \\
\textbf{Ca I   } &  0.08 &  0.02 &  0.04 &  0.01 &   0.03 &      0.08 &        0.10 &       0.12 \\
\textbf{Sc I   } &  0.12 &  0.01 &  0.02 &  0.02 &  —     &      0.20 &          —  &       0.24 \\
\textbf{Ti I   } &  0.19 &  0.00 &  0.06 &  0.01 &   0.08 &      0.09 &        0.21 &       0.22 \\
\textbf{Ti II  } &  0.00 &  0.03 &  0.01 &  0.01 &   0.09 &      0.09 &        0.09 &       0.10 \\
\textbf{Cr I   } &  0.16 &  0.00 &  0.06 &  0.02 &   0.11 &      0.09 &        0.21 &       0.19 \\
\textbf{Ni I   } &  0.14 &  0.00 &  0.04 &  0.01 &   0.14 &      0.12 &        0.20 &       0.19 \\
\textbf{Ba II  } &  0.05 &  0.04 &  0.02 &  0.01 &   0.14 &      0.13 &        0.15 &       0.14 \\
\midrule
\textbf{CB-2}    &       &       &       &       &        &           &             &       \\
\textbf{Fe I   } &  0.10 &  0.01 &  0.03 &  0.02 &   0.03 &      0.05 &        0.12 &       0.12 \\
\textbf{Fe II  } &  0.00 &  0.05 &  0.02 &  0.01 &   0.10 &      0.14 &        0.11 &       0.15 \\
\textbf{Na I   } &  0.11 &  0.02 &  0.08 &  0.05 &   0.16 &      0.29 &        0.22 &       0.33 \\
\textbf{Mg I   } &  0.09 &  0.03 &  0.04 &  0.04 &   0.13 &      0.25 &        0.17 &       0.28 \\
\textbf{\ion{K}{I}}& —   &  —    &  —    &  —    &   —    &      —    &        —    &       —    \\
\textbf{Ca I   } &  0.07 &  0.01 &  0.01 &  0.01 &   0.10 &      0.13 &        0.12 &       0.15 \\
\textbf{Sc I   } &  0.09 &  0.01 &  0.02 &  0.01 &      — &      0.20 &           — &       0.22 \\
\textbf{Ti I   } &  0.12 &  0.01 &  0.04 &  0.02 &   0.10 &      0.15 &        0.16 &       0.20 \\
\textbf{Ti II  } &  0.03 &  0.05 &  0.02 &  0.02 &   0.23 &      0.41 &        0.23 &       0.41 \\
\textbf{Cr I   } &  0.12 &  0.02 &  0.04 &  0.03 &   0.13 &      0.27 &        0.18 &       0.30 \\
\textbf{Ni I   } &  0.12 &  0.01 &  0.03 &  0.02 &   0.10 &      0.11 &        0.16 &       0.17 \\
\textbf{Ba II  } &  0.07 &  0.06 &  0.03 &  0.00 &   0.16 &      0.17 &        0.18 &       0.19 \\
\midrule
\textbf{UMaI-1}    &       &       &       &       &        &           &             &       \\
\textbf{Fe I   } &  0.11 &  0.01 &  0.02 &  0.02 &   0.03 &      0.07 &        0.12 &       0.13 \\
\textbf{Fe II  } &  0.05 &  0.04 &  0.02 &  0.01 &   0.13 &      0.16 &        0.15 &       0.17 \\
\textbf{Na I   } &  0.14 &  0.01 &  0.03 &  0.04 &   0.13 &      0.61 &        0.20 &       0.63 \\
\textbf{Mg I   } &  0.09 &  0.02 &  0.02 &  0.02 &   0.11 &      0.34 &        0.14 &       0.35 \\
\textbf{\ion{K}{I}}&0.12 &  0.02 &  0.03 &  0.01 &   0.18 &      —    &        0.22 &           — \\
\textbf{Ca I   } &  0.09 &  0.00 &  0.02 &  0.01 &   0.10 &      0.27 &        0.14 &       0.29 \\
\textbf{Sc I   } &  0.12 &  0.00 &  0.03 &  0.01 &      — &      0.20 &          —  &       0.24 \\
\textbf{Ti I   } &  0.19 &  0.03 &  0.04 &  0.04 &   0.13 &      1.01 &        0.24 &       1.03 \\
\textbf{Ti II  } &  0.01 &  0.04 &  0.01 &  0.03 &   0.11 &      0.33 &        0.12 &       0.33 \\
\textbf{Cr I   } &  0.15 &  0.01 &  0.04 &  0.01 &   0.13 &       0.4 &        0.21 &       0.43 \\
\textbf{Ni I   } &  0.10 &  0.01 &  0.01 &  0.02 &   0.09 &      0.34 &        0.14 &       0.36 \\
\textbf{Ba II  } &  0.02 &  0.05 &  0.02 &  0.02 &   0.11 &      0.31 &        0.12 &       0.32 \\
\midrule
\textbf{UMaI-2}    &       &       &       &       &        &           &             &       \\
\textbf{Fe I   } &  0.19 &  0.03 &  0.01 &  0.02 &   0.04 &      0.06 &        0.20 &       0.20 \\
\textbf{Fe II  } &  0.03 &  0.01 &  0.00 &  0.03 &   0.16 &      0.23 &        0.17 &       0.23 \\
\textbf{Na I   } &  0.21 &  0.04 &  0.01 &  0.02 &   0.20 &      0.38 &        0.29 &       0.44 \\
\textbf{Mg I   } &  0.16 &  0.02 &  0.01 &  0.01 &   0.16 &      0.36 &        0.23 &       0.40 \\
\textbf{\ion{K}{I}}&0.12 &  0.02 &  0.01 &  0.01 &   0.28 &      —    &        0.30 &       —    \\
\textbf{Ca I   } &  0.12 &  0.02 &  0.01 &  0.01 &   0.11 &      0.17 &        0.16 &       0.21 \\
\textbf{Sc I   } &  0.14 &  0.02 &  0.01 &  0.02 &      — &      0.20 &           — &       0.24 \\
\textbf{\ion{Ti}{I}}& —   &  —    &  —    &  —    &   —    &      —    &        —    &       —    \\
\textbf{\ion{Ti}{II}}& —   &  —    &  —    &  —    &   —    &      —    &        —    &       —    \\
\textbf{Cr I   } &  0.19 &  0.03 &  0.01 &  0.02 &   0.20 &      0.29 &        0.27 &       0.35 \\
\textbf{Ni I   } &  0.18 &  0.03 &  0.01 &  0.02 &   0.20 &      0.19 &        0.27 &       0.27 \\
\textbf{Ba II  } &  0.03 &  0.02 &  0.01 &  0.04 &   0.16 &      0.16 &        0.17 &       0.17 \\

\bottomrule
\end{tabular}
\label{tab:errorsfull}
\end{table}

\begin{table}
\ContinuedFloat
\caption{}
\begin{tabular}{lrrrrrrrr}
\toprule
\textbf{species} & $\Delta T$ &   $\Delta g$ &   $\Delta m$ &   $\Delta \xi$ &  $\sigma_{EW}$ & $\sigma_{syn}$ &  $\Delta$EW &  $\Delta$syn \\
\midrule
\textbf{BooI-2}    &       &       &       &       &        &           &             &       \\
\textbf{Fe I   } &  0.12 &  0.00 &  0.06 &  0.02 &   0.03 &      0.06 &        0.14 &       0.15 \\
\textbf{Fe II  } &  0.04 &  0.02 &  0.01 &  0.02 &   0.08 &       0.1 &        0.10 &       0.11 \\
\textbf{Na I   } &  0.15 &  0.01 &  0.07 &  0.03 &   0.15 &      0.21 &        0.22 &       0.27 \\
\textbf{Mg I   } &  0.07 &  0.01 &  0.06 &  0.02 &   0.21 &      0.22 &        0.23 &       0.24 \\
\textbf{\ion{K}{I}}&0.11 &  0.00 &  0.05 &  0.01 &   0.21 &      —    &        0.24 &       —    \\
\textbf{Ca I   } &  0.08 &  0.01 &  0.04 &  0.01 &   0.06 &      0.12 &        0.11 &       0.15 \\
\textbf{Sc I   } &  0.12 &  0.00 &  0.02 &  0.02 &      — &      0.20 &           — &       0.24 \\
\textbf{Ti I   } &  0.17 &  0.00 &  0.08 &  0.03 &   0.10 &       0.2 &        0.21 &       0.27 \\
\textbf{Ti II  } &  0.01 &  0.02 &  0.03 &  0.01 &   0.15 &      0.36 &        0.15 &       0.36 \\
\textbf{Cr I   } &  0.14 &  0.00 &  0.06 &  0.01 &   0.15 &      0.17 &        0.21 &       0.23 \\
\textbf{Ni I   } &  0.12 &  0.00 &  0.04 &  0.01 &   0.09 &      0.13 &        0.16 &       0.18 \\
\textbf{Ba II  } &  0.03 &  0.03 &  0.04 &  0.01 &   0.15 &      0.16 &        0.16 &       0.17 \\
\bottomrule
\end{tabular}
\end{table}

\bsp	
\label{lastpage}
\end{document}